\documentclass[a4paper,11pt]{article}
\usepackage{jcappub} 
\usepackage{bm,physics,upgreek} 

\usepackage[output-decimal-marker={.}, separate-uncertainty=true]{siunitx} 
\AtBeginDocument{\RenewCommandCopy\qty\SI}

\usepackage[english]{babel}
\usepackage{csquotes}
\usepackage{xcolor}

\usepackage{lineno}
\usepackage[noabbrev]{cleveref} 
\crefname{equation}{eq.}{eqs.}
\Crefname{equation}{Eq.}{Eqs.}

\usepackage{float}
\usepackage{comment}
\usepackage{booktabs,microtype,multirow}
\usepackage{color,soul}

\definecolor{tabblue}{HTML}{1F77B4}

\definecolor{tabgreen}{HTML}{1EC213}

\newcommand{\Expval}{\mathbb{E}}

\newcommand{\Nsample}{\num{10000}}

\newcommand{\Eth}{E_\text{th}}

\newcommand{\background}{\text{bg}}
\newcommand{\source}{\text{source}}

\newcommand{\ee}{\text{e}}

\newcommand{\cut}{\text{cut}}

\newcommand{\on}{\text{on}}
\newcommand{\off}{\text{off}}
\newcommand{\tot}{\text{tot}}

\newcommand{\tmin}{\text{min}}
\newcommand{\tmax}{\text{max}}
\newcommand{\Emax}{E_\tmax}
\newcommand{\Emin}{E_\tmin}
\newcommand{\ts}{\text{s}}

\newcommand{\LL}{\mathcal{L}}

\newcommand{\EE}{\mathcal{E}}

\newcommand{\tH}{\text{H}}

\newcommand{\TS}{\text{TS}}

\DeclareSIUnit \parsec {pc}
\DeclareSIUnit{\pc}{\parsec}
\DeclareSIUnit\kpc {\kilo\pc}
\DeclareSIUnit \Mpc {Mpc}
\DeclareSIUnit\Gpc{\giga\parsec}
\DeclareSIUnit{\gauss}{G}
\DeclareSIUnit{\muG}{\micro\gauss}
\DeclareSIUnit{\mG}{\milli\gauss}

\DeclareSIUnit \PeV {PeV} 
\DeclareSIUnit \EeV {EeV} 
\DeclareSIUnit \VEM {VEM} 
\DeclareSIUnit \erg {erg}
\DeclareSIUnit \s {s}
\DeclareSIUnit{\yr}{yr}
\DeclareSIUnit{\day}{day}
\DeclareSIUnit\mag {mag}
\DeclareSIUnit \Jy {Jy}

\DeclareSIUnit{\h}{\hour}
\DeclareSIUnit{\sr}{sr}

\title{Hidden structures in the UHE sky: unveiling potential cosmic-ray sources via spectral signatures}

\author[a]{M.\ Z.\ Rennó}
\author[b]{and C.\ de Oliveira}

\affiliation[a]{Universidade Estadual de Campinas, IFGW, Campinas, SP, Brazil}
\affiliation[b]{Universidade de São Paulo, Instituto de Física, São Paulo, SP, Brazil}
\emailAdd{renno.mateus@dac.unicamp.br, caina.oliveira@usp.br}

\abstract{
        The search for the origin of ultrahigh-energy cosmic rays (UHECRs) typically relies on identifying directional excesses in the particle flux.
        While this approach is well established, it cannot detect localized spectral features, which could provide constraints on the sources of UHECRs if measured.
        In this work, we propose extending directional searches through two complementary energy-weighted metrics: the accumulated energy $U$ and the local average energy per particle $\mu_E$. We demonstrate that $U$ inherently amplifies the statistical significance of sources with high-energy tails, while $\mu_E$ provides a distinct probe for direction-dependent spectra.
        To assess the statistical significance of these metrics, we derive a generalization of the Li–Ma significance formula for continuous energy variables. 
        Using a maximum-likelihood analysis within an exploratory Monte-Carlo model, we show that these variables can enhance discovery potential by amplifying  localized excesses and identifying hard-spectrum regions that are not apparent in standard particle-flux maps, thereby providing a comprehensive probe of potential UHECR sources. 
        Furthermore, $\mu_E$ variations associated with localized excesses are smoothed into minor perturbations when the spectrum is integrated over right-ascension or declination bands.
        Consequently, the presence of small- or intermediate-scale regions with distinct spectral shapes remains consistent with the absence of declination-band spectral variations recently reported by the Pierre Auger Collaboration.

}

\begin{document}
\maketitle
\flushbottom
\newcommand{\PrintTabResultsDec}{
  \resizebox{\textwidth}{!}{%
  \begin{tabular}{ccccccc}
    \toprule
    $\Eth$ & $0^\circ \text{ to } 60^\circ$** & $60^\circ \text{ to } 120^\circ$ & $120^\circ \text{ to } 180^\circ$ & $180^\circ \text{ to } 240^\circ$* & $240^\circ \text{ to } 300^\circ$ & $300^\circ \text{ to } 360^\circ$ \\
    \midrule
    $\qty{10}{\EeV}$ & $\qty{18.0 \pm 0.2}{\EeV}$ & $\qty{17.7 \pm 0.1}{\EeV}$ & $\qty{17.8 \pm 0.1}{\EeV}$ & $\qty{18.0 \pm 0.1}{\EeV}$ & $\qty{17.7 \pm 0.1}{\EeV}$ & $\qty{17.8 \pm 0.1}{\EeV}$ \\
    $\qty{20}{\EeV}$ & $\qty{32.1 \pm 0.4}{\EeV}$ & $\qty{31.2 \pm 0.3}{\EeV}$ & $\qty{31.3 \pm 0.3}{\EeV}$ & $\qty{31.6 \pm 0.3}{\EeV}$ & $\qty{31.2 \pm 0.3}{\EeV}$ & $\qty{31.4 \pm 0.3}{\EeV}$ \\
    $\qty{40}{\EeV}$ & $\qty{53.7 \pm 0.9}{\EeV}$ & $\qty{53 \pm 1}{\EeV}$ & $\qty{53.0 \pm 0.9}{\EeV}$ & $\qty{53.3 \pm 0.9}{\EeV}$ & $\qty{53 \pm 1}{\EeV}$ & $\qty{53 \pm 1}{\EeV}$ \\
    $\qty{50}{\EeV}$ & $\qty{64 \pm 1}{\EeV}$ & $\qty{64 \pm 2}{\EeV}$ & $\qty{64 \pm 2}{\EeV}$ & $\qty{65 \pm 2}{\EeV}$ & $\qty{64 \pm 2}{\EeV}$ & $\qty{64 \pm 2}{\EeV}$ \\
    \bottomrule
  \end{tabular} } }

\newcommand{\PrintTabResultsRA}{
  \resizebox{\textwidth}{!}{%
  \begin{tabular}{cccccc}
    \toprule
    $\Eth$ & $-90.0^\circ \text{ to } -51.0^\circ$ & $-51.0^\circ \text{ to } -29.0^\circ$* & $-29.0^\circ \text{ to } -8.0^\circ$ & $-8.0^\circ \text{ to } 24.8^\circ$ & $24.8^\circ \text{ to } 44.8^\circ$** \\
    \midrule
    $\qty{10}{\EeV}$ & $\qty{17.8 \pm 0.1}{\EeV}$ & $\qty{17.86 \pm 0.08}{\EeV}$ & $\qty{17.75 \pm 0.08}{\EeV}$ & $\qty{17.75 \pm 0.09}{\EeV}$ & $\qty{18.0 \pm 0.2}{\EeV}$ \\
    $\qty{20}{\EeV}$ & $\qty{31.4 \pm 0.2}{\EeV}$ & $\qty{31.4 \pm 0.2}{\EeV}$ & $\qty{31.3 \pm 0.2}{\EeV}$ & $\qty{31.3 \pm 0.2}{\EeV}$ & $\qty{32.0 \pm 0.5}{\EeV}$ \\
    $\qty{40}{\EeV}$ & $\qty{53.1 \pm 0.7}{\EeV}$ & $\qty{53.1 \pm 0.5}{\EeV}$ & $\qty{53.0 \pm 0.6}{\EeV}$ & $\qty{53.0 \pm 0.6}{\EeV}$ & $\qty{54 \pm 1}{\EeV}$ \\
    $\qty{50}{\EeV}$ & $\qty{64 \pm 1}{\EeV}$ & $\qty{64.5 \pm 0.9}{\EeV}$ & $\qty{64 \pm 1}{\EeV}$ & $\qty{64 \pm 1}{\EeV}$ & $\qty{64 \pm 2}{\EeV}$ \\
    \bottomrule
  \end{tabular} } }

\newcommand{\PrintTabResultsTS}{
  \resizebox{\textwidth}{!}{%
  \begin{tabular}{ccccccccc}
    \toprule
    Metric & $E_{\text{th}}$ (\unit{\EeV}) & $\Psi$ ($^\circ$) & TS & $S$ & $\ell$ ($^\circ$) & $b$ ($^\circ$) & Expected & Observed \\
    \midrule
    \multicolumn{9}{c}{\textbf{Search in the full field of view}} \\
    \midrule
  $N$            & $20$ & $27^{+1}_{-1}$ & $29^{+9}_{-9}$ & $5.3^{+0.9}_{-0.9}$ & $309^{+3}_{-3}$ & $17^{+3}_{-3}$ & $828 \pm 30$ & $995 \pm 42$ \\
                 & $40$ & $27^{+1}_{-1}$ & $29^{+8}_{-8}$ & $5.4^{+0.8}_{-0.8}$ & $309^{+3}_{-3}$ & $17^{+3}_{-3}$ & $144 \pm 6$ & $217 \pm 13$ \\
    \midrule
  $U$            & $20$ & $27^{+1}_{-1}$ & $38^{+11}_{-11}$ & $6.1^{+0.9}_{-0.9}$ & $309^{+2}_{-3}$ & $17^{+2}_{-2}$ & $(25900 \pm 800)\,\unit{\EeV}$ & $(32000 \pm 1000)\,\unit{\EeV}$ \\
                 & $50$ & $20^{+1}_{-1}$ & $47^{+20}_{-20}$ & $6.7^{+1.5}_{-1.5}$ & $127^{+3}_{-3}$ & $-25^{+4}_{-3}$ & $(330 \pm 90)\,\unit{\EeV}$ & $(1300 \pm 300)\,\unit{\EeV}$ \\
    \midrule
  $\hat\mu_E$    & $10$ & $20^{+1}_{-1}$ & $38^{+13}_{-13}$ & $6.1^{+1.1}_{-1.1}$ & $127^{+3}_{-3}$ & $-26^{+3}_{-3}$ & $(17.78 \pm 0.05)\,\unit{\EeV}$ & $(21.5 \pm 0.8)\,\unit{\EeV}$ \\
    \midrule
    \multicolumn{9}{c}{\textbf{Search excluding first-peak region}} \\
    \midrule
  $N$            & $20$ & $20^{+1}_{-1}$ & $15^{+4}_{-4}$ & $3.8^{+0.5}_{-0.5}$ & $127^{+7}_{-7}$ & $-27^{+6}_{-6}$ & $75 \pm 29$ & $110 \pm 35$ \\
                 & $40$ & $22^{+1}_{-2}$ & $19^{+5}_{-5}$ & $4.4^{+0.6}_{-0.6}$ & $127^{+5}_{-5}$ & $-27^{+5}_{-5}$ & $16 \pm 7$ & $36 \pm 11$ \\
    \midrule
                 & $20$ & $20^{+1}_{-1}$ & $23^{+6}_{-6}$ & $4.8^{+0.7}_{-0.7}$ & $127^{+5}_{-5}$ & $-26^{+5}_{-3}$ & $(2200 \pm 800)\,\unit{\EeV}$ & $(4000 \pm 1000)\,\unit{\EeV}$ \\
  $U$            & $40$ & $25^{+1}_{-3}$ & $26^{+8}_{-8}$ & $5.0^{+0.8}_{-0.8}$ & $309^{+5}_{-5}$ & $17^{+5}_{-5}$ & $(6600 \pm 400)\,\unit{\EeV}$ & $(9900 \pm 700)\,\unit{\EeV}$ \\
    \midrule
  $\hat\mu_E$    & $10$ & $24^{+1}_{-2}$ & $23^{+7}_{-7}$ & $4.7^{+0.8}_{-0.8}$ & $309^{+6}_{-6}$ & $17^{+6}_{-6}$ & $(17.73 \pm 0.05)\,\unit{\EeV}$ & $(18.7 \pm 0.2)\,\unit{\EeV}$ \\
    \bottomrule
  \end{tabular} } } 
\newcommand{\PrintTabGlobalN}{
  \resizebox{\textwidth}{!}{%
  \begin{tabular}{ccccccccc}
    \toprule
    $E_{\text{th}}$ (\unit{\EeV}) & $\Psi$ ($^\circ$) & TS & $S$ & $\ell$ ($^\circ$) & $b$ ($^\circ$) & $N_{\text{exp}}$ & $N_{\text{obs}}$ & Probable source \\
    \midrule
    $8$ & $20^{+1}_{-1}$ & $13^{+4}_{-4}$ & $3.5^{+0.5}_{-0.5}$ & $309^{+10}_{-10}$ & $16^{+9}_{-9}$ & $2870 \pm 190$ & $3060 \pm 200$ & Cen~A \\
    $10$ & $22^{+1}_{-2}$ & $14^{+4}_{-4}$ & $3.7^{+0.6}_{-0.6}$ & $309^{+9}_{-9}$ & $17^{+8}_{-8}$ & $2310 \pm 150$ & $2500 \pm 160$ & Cen~A \\
    $13$ & $25^{+1}_{-1}$ & $17^{+6}_{-6}$ & $4.0^{+0.7}_{-0.7}$ & $309^{+7}_{-6}$ & $17^{+7}_{-6}$ & $1780 \pm 100$ & $1960 \pm 110$ & Cen~A \\
    $16$ & $27^{+1}_{-1}$ & $21^{+7}_{-7}$ & $4.5^{+0.8}_{-0.8}$ & $309^{+5}_{-5}$ & $17^{+5}_{-5}$ & $1337 \pm 65$ & $1514 \pm 76$ & Cen~A \\
    $20$ & $27^{+1}_{-1}$ & $29^{+9}_{-9}$ & $5.3^{+0.9}_{-0.9}$ & $309^{+3}_{-3}$ & $17^{+3}_{-3}$ & $828 \pm 30$ & $995 \pm 42$ & Cen~A \\
    $25$ & $25^{+1}_{-1}$ & $21^{+7}_{-7}$ & $4.5^{+0.7}_{-0.7}$ & $309^{+5}_{-5}$ & $17^{+5}_{-5}$ & $431 \pm 22$ & $533 \pm 29$ & Cen~A \\
    $32$ & $26^{+1}_{-1}$ & $27^{+8}_{-8}$ & $5.1^{+0.8}_{-0.8}$ & $309^{+4}_{-4}$ & $17^{+4}_{-3}$ & $254 \pm 11$ & $344 \pm 19$ & Cen~A \\
    $40$ & $27^{+1}_{-1}$ & $29^{+8}_{-8}$ & $5.4^{+0.8}_{-0.8}$ & $309^{+3}_{-3}$ & $17^{+3}_{-3}$ & $144 \pm 6$ & $217 \pm 13$ & Cen~A \\
    $50$ & $20^{+1}_{-1}$ & $29^{+10}_{-10}$ & $5.3^{+0.9}_{-0.9}$ & $127^{+3}_{-3}$ & $-27^{+3}_{-2}$ & $6 \pm 1$ & $22 \pm 4$ & PPSC \\
    $63$ & $20^{+1}_{-1}$ & $19^{+7}_{-7}$ & $4.3^{+0.8}_{-0.8}$ & $127^{+5}_{-5}$ & $-27^{+5}_{-4}$ & $2 \pm 1$ & $11 \pm 3$ & PPSC \\
    $79$ & $20^{+1}_{-1}$ & $12^{+4}_{-4}$ & $3.4^{+0.5}_{-0.5}$ & $308^{+10}_{-10}$ & $17^{+9}_{-9}$ & $4 \pm 1$ & $14 \pm 2$ & Cen~A \\
    \bottomrule
  \end{tabular}%
 } }
\newcommand{\PrintTabLocalN}{
  \resizebox{\textwidth}{!}{%
  \begin{tabular}{ccccccccc}
    \toprule
    $E_{\text{th}}$ (\unit{\EeV}) & $\Psi$ ($^\circ$) & TS & $S$ & $\ell$ ($^\circ$) & $b$ ($^\circ$) & $N_{\text{exp}}$ & $N_{\text{obs}}$ & Probable source \\
    \midrule
    $8$ & $20^{+1}_{-1}$ & $9^{+2}_{-2}$ & $2.9^{+0.3}_{-0.4}$ & $183^{+12}_{-10}$ & $4^{+11}_{-12}$ & $700 \pm 330$ & $770 \pm 350$ & Random \\
    $10$ & $20^{+1}_{-1}$ & $9^{+2}_{-2}$ & $2.9^{+0.3}_{-0.3}$ & $128^{+13}_{-13}$ & $-30^{+11}_{-11}$ & $390 \pm 200$ & $440 \pm 210$ & PPSC \\
    $13$ & $20^{+2}_{-1}$ & $10^{+2}_{-3}$ & $3.1^{+0.4}_{-0.4}$ & $127^{+12}_{-11}$ & $-28^{+8}_{-9}$ & $207 \pm 98$ & $250 \pm 110$ & PPSC \\
    $16$ & $20^{+1}_{-1}$ & $12^{+3}_{-3}$ & $3.4^{+0.4}_{-0.4}$ & $127^{+10}_{-9}$ & $-28^{+7}_{-7}$ & $129 \pm 58$ & $169 \pm 67$ & PPSC \\
    $20$ & $20^{+1}_{-1}$ & $15^{+4}_{-4}$ & $3.8^{+0.5}_{-0.5}$ & $127^{+7}_{-7}$ & $-27^{+6}_{-6}$ & $75 \pm 29$ & $110 \pm 35$ & PPSC \\
    $25$ & $20^{+1}_{-1}$ & $14^{+4}_{-4}$ & $3.7^{+0.5}_{-0.5}$ & $127^{+7}_{-6}$ & $-28^{+6}_{-6}$ & $48 \pm 20$ & $76 \pm 25$ & PPSC \\
    $32$ & $20^{+1}_{-1}$ & $17^{+5}_{-5}$ & $4.1^{+0.6}_{-0.6}$ & $127^{+5}_{-5}$ & $-28^{+5}_{-4}$ & $26 \pm 10$ & $50 \pm 14$ & PPSC \\
    $40$ & $22^{+1}_{-2}$ & $19^{+5}_{-5}$ & $4.4^{+0.6}_{-0.6}$ & $127^{+5}_{-5}$ & $-27^{+5}_{-5}$ & $16 \pm 7$ & $36 \pm 11$ & PPSC \\
    $50$ & $22^{+1}_{-1}$ & $18^{+5}_{-5}$ & $4.1^{+0.6}_{-0.6}$ & $309^{+8}_{-8}$ & $17^{+7}_{-7}$ & $42 \pm 3$ & $73 \pm 6$ & Cen~A \\
    $63$ & $20^{+1}_{-1}$ & $12^{+4}_{-4}$ & $3.4^{+0.5}_{-0.5}$ & $309^{+10}_{-10}$ & $17^{+9}_{-9}$ & $13 \pm 1$ & $27 \pm 3$ & Cen~A \\
    $79$ & $20^{+1}_{-1}$ & $9^{+2}_{-2}$ & $2.9^{+0.4}_{-0.4}$ & $125^{+11}_{-11}$ & $-29^{+9}_{-11}$ & $1 \pm 1$ & $5 \pm 1$ & PPSC \\
    \bottomrule
  \end{tabular}%
 } }
\newcommand{\PrintTabGlobalU}{
  \resizebox{\textwidth}{!}{%
  \begin{tabular}{ccccccccc}
    \toprule
    $E_{\text{th}}$ (\unit{\EeV}) & $\Psi$ ($^\circ$) & TS & $S$ & $\ell$ ($^\circ$) & $b$ ($^\circ$) & $U_{\text{exp}}$ (\unit{\EeV}) & $U_{\text{obs}}$ (\unit{\EeV}) & Probable source \\
    \midrule
    $8$ & $26^{+1}_{-1}$ & $26^{+8}_{-8}$ & $5.0^{+0.8}_{-0.8}$ & $309^{+4}_{-4}$ & $17^{+4}_{-4}$ & $71000 \pm 3000$ & $77000 \pm 3000$ & Cen~A \\
    $10$ & $26^{+1}_{-1}$ & $27^{+9}_{-9}$ & $5.2^{+0.9}_{-0.9}$ & $309^{+4}_{-4}$ & $17^{+4}_{-4}$ & $57000 \pm 2000$ & $63000 \pm 3000$ & Cen~A \\
    $13$ & $27^{+1}_{-1}$ & $30^{+9}_{-10}$ & $5.4^{+0.9}_{-0.9}$ & $309^{+3}_{-3}$ & $17^{+3}_{-3}$ & $45000 \pm 2000$ & $52000 \pm 2000$ & Cen~A \\
    $16$ & $27^{+1}_{-1}$ & $34^{+10}_{-10}$ & $5.7^{+0.9}_{-0.9}$ & $309^{+3}_{-3}$ & $17^{+3}_{-3}$ & $35000 \pm 1000$ & $41000 \pm 2000$ & Cen~A \\
    $20$ & $27^{+1}_{-1}$ & $38^{+11}_{-11}$ & $6.1^{+0.9}_{-0.9}$ & $309^{+2}_{-3}$ & $17^{+2}_{-2}$ & $25900 \pm 800$ & $32000 \pm 1000$ & Cen~A \\
    $25$ & $20^{+1}_{-1}$ & $36^{+13}_{-12}$ & $5.9^{+1.1}_{-1.0}$ & $127^{+3}_{-3}$ & $-26^{+3}_{-3}$ & $1500 \pm 400$ & $3000 \pm 600$ & PPSC \\
    $32$ & $20^{+1}_{-1}$ & $42^{+15}_{-14}$ & $6.4^{+1.2}_{-1.1}$ & $127^{+3}_{-3}$ & $-26^{+3}_{-2}$ & $1000 \pm 300$ & $2400 \pm 500$ & PPSC \\
    $40$ & $20^{+1}_{-1}$ & $46^{+17}_{-17}$ & $6.7^{+1.3}_{-1.2}$ & $127^{+3}_{-3}$ & $-26^{+3}_{-2}$ & $600 \pm 200$ & $1900 \pm 400$ & PPSC \\
    $50$ & $20^{+1}_{-1}$ & $47^{+20}_{-20}$ & $6.7^{+1.5}_{-1.5}$ & $127^{+3}_{-3}$ & $-25^{+4}_{-3}$ & $330 \pm 90$ & $1300 \pm 300$ & PPSC \\
    $63$ & $20^{+1}_{-1}$ & $32^{+14}_{-14}$ & $5.5^{+1.3}_{-1.3}$ & $127^{+6}_{-6}$ & $-25^{+6}_{-4}$ & $140 \pm 50$ & $700 \pm 200$ & PPSC \\
    $79$ & $20^{+1}_{-1}$ & $19^{+6}_{-7}$ & $4.3^{+0.7}_{-0.8}$ & $309^{+10}_{-10}$ & $17^{+9}_{-9}$ & $430 \pm 60$ & $1400 \pm 200$ & Cen~A \\
    \bottomrule
  \end{tabular}%
 } }
\newcommand{\PrintTabLocalU}{
  \resizebox{\textwidth}{!}{%
  \begin{tabular}{ccccccccc}
    \toprule
    $E_{\text{th}}$ (\unit{\EeV}) & $\Psi$ ($^\circ$) & TS & $S$ & $\ell$ ($^\circ$) & $b$ ($^\circ$) & $U_{\text{exp}}$ (\unit{\EeV}) & $U_{\text{obs}}$ (\unit{\EeV}) & Probable source \\
    \midrule
    $8$ & $20^{+1}_{-1}$ & $15^{+4}_{-4}$ & $3.9^{+0.6}_{-0.6}$ & $127^{+7}_{-7}$ & $-26^{+6}_{-5}$ & $7000 \pm 3000$ & $8000 \pm 3000$ & PPSC \\
    $10$ & $20^{+1}_{-1}$ & $16^{+5}_{-5}$ & $4.0^{+0.6}_{-0.6}$ & $128^{+7}_{-7}$ & $-26^{+6}_{-5}$ & $5000 \pm 2000$ & $7000 \pm 2000$ & PPSC \\
    $13$ & $20^{+1}_{-1}$ & $18^{+5}_{-5}$ & $4.2^{+0.6}_{-0.6}$ & $127^{+6}_{-6}$ & $-26^{+5}_{-4}$ & $4000 \pm 1000$ & $5000 \pm 2000$ & PPSC \\
    $16$ & $20^{+1}_{-1}$ & $20^{+6}_{-6}$ & $4.5^{+0.7}_{-0.6}$ & $127^{+5}_{-5}$ & $-26^{+5}_{-4}$ & $3000 \pm 1000$ & $4000 \pm 1000$ & PPSC \\
    $20$ & $20^{+1}_{-1}$ & $23^{+6}_{-6}$ & $4.8^{+0.7}_{-0.7}$ & $127^{+5}_{-5}$ & $-26^{+5}_{-3}$ & $2200 \pm 800$ & $4000 \pm 1000$ & PPSC \\
    $25$ & $20^{+1}_{-1}$ & $22^{+7}_{-7}$ & $4.6^{+0.7}_{-0.7}$ & $309^{+9}_{-9}$ & $16^{+9}_{-8}$ & $10500 \pm 700$ & $13600 \pm 900$ & Cen~A \\
    $32$ & $23^{+2}_{-1}$ & $25^{+8}_{-8}$ & $4.9^{+0.8}_{-0.8}$ & $309^{+7}_{-7}$ & $17^{+6}_{-6}$ & $9000 \pm 600$ & $12500 \pm 900$ & Cen~A \\
    $40$ & $25^{+1}_{-3}$ & $26^{+8}_{-8}$ & $5.0^{+0.8}_{-0.8}$ & $309^{+5}_{-5}$ & $17^{+5}_{-5}$ & $6600 \pm 400$ & $9900 \pm 700$ & Cen~A \\
    $50$ & $20^{+1}_{-1}$ & $23^{+7}_{-7}$ & $4.7^{+0.8}_{-0.8}$ & $309^{+9}_{-8}$ & $17^{+8}_{-8}$ & $2200 \pm 200$ & $4200 \pm 400$ & Cen~A \\
    $63$ & $20^{+1}_{-1}$ & $17^{+5}_{-5}$ & $4.0^{+0.7}_{-0.7}$ & $309^{+10}_{-10}$ & $17^{+9}_{-9}$ & $1000 \pm 90$ & $2200 \pm 300$ & Cen~A \\
    $79$ & $20^{+1}_{-1}$ & $15^{+5}_{-5}$ & $3.9^{+0.6}_{-0.7}$ & $128^{+12}_{-11}$ & $-27^{+9}_{-10}$ & $80 \pm 40$ & $400 \pm 100$ & PPSC \\
    \bottomrule
  \end{tabular}%
 } }
\newcommand{\PrintTabGlobalE}{
  \resizebox{\textwidth}{!}{%
  \begin{tabular}{ccccccccc}
    \toprule
    $E_{\text{th}}$ (\unit{\EeV}) & $\Psi$ ($^\circ$) & TS & $S$ & $\ell$ ($^\circ$) & $b$ ($^\circ$) & $\mu_{E, \text{exp}}$ (\unit{\EeV}) & $\hat{\mu}_{E, \text{obs}}$ (\unit{\EeV}) & Probable source \\
    \midrule
    $8$ & $20^{+1}_{-1}$ & $36^{+12}_{-12}$ & $5.9^{+1.0}_{-1.0}$ & $127^{+3}_{-3}$ & $-26^{+3}_{-3}$ & $14.86 \pm 0.04$ & $17.7 \pm 0.6$ & PPSC \\
    $10$ & $20^{+1}_{-1}$ & $38^{+13}_{-13}$ & $6.1^{+1.1}_{-1.1}$ & $127^{+3}_{-3}$ & $-26^{+3}_{-3}$ & $17.78 \pm 0.05$ & $21.5 \pm 0.8$ & PPSC \\
    $13$ & $20^{+1}_{-1}$ & $37^{+13}_{-13}$ & $6.0^{+1.1}_{-1.1}$ & $127^{+3}_{-3}$ & $-26^{+3}_{-2}$ & $22.05 \pm 0.08$ & $27 \pm 1$ & PPSC \\
    $16$ & $20^{+1}_{-1}$ & $35^{+13}_{-13}$ & $5.8^{+1.1}_{-1.1}$ & $127^{+3}_{-3}$ & $-26^{+3}_{-2}$ & $26.2 \pm 0.1$ & $33 \pm 1$ & PPSC \\
    $20$ & $20^{+1}_{-1}$ & $29^{+11}_{-11}$ & $5.3^{+1.0}_{-1.0}$ & $127^{+4}_{-4}$ & $-27^{+4}_{-3}$ & $31.3 \pm 0.1$ & $39 \pm 2$ & PPSC \\
    $25$ & $20^{+1}_{-1}$ & $24^{+8}_{-8}$ & $4.8^{+0.9}_{-0.9}$ & $127^{+5}_{-5}$ & $-27^{+5}_{-4}$ & $37.3 \pm 0.2$ & $46 \pm 2$ & PPSC \\
    $32$ & $20^{+1}_{-1}$ & $18^{+6}_{-6}$ & $4.2^{+0.7}_{-0.7}$ & $127^{+8}_{-8}$ & $-28^{+7}_{-7}$ & $44.8 \pm 0.3$ & $54 \pm 3$ & PPSC \\
    $40$ & $20^{+1}_{-1}$ & $15^{+5}_{-5}$ & $3.8^{+0.6}_{-0.6}$ & $307^{+11}_{-11}$ & $15^{+10}_{-10}$ & $52.7 \pm 0.4$ & $58 \pm 1$ & Cen~A \\
    $50$ & $20^{+1}_{-1}$ & $15^{+4}_{-5}$ & $3.8^{+0.6}_{-0.6}$ & $309^{+12}_{-12}$ & $18^{+10}_{-10}$ & $63.7 \pm 0.6$ & $72 \pm 2$ & Cen~A \\
    $63$ & $20^{+1}_{-1}$ & $16^{+6}_{-6}$ & $4.0^{+0.8}_{-0.8}$ & $273^{+11}_{-11}$ & $-13^{+11}_{-10}$ & $79 \pm 1$ & $100 \pm 7$ & Random \\
    $79$ & $20^{+1}_{-1}$ & $15^{+6}_{-6}$ & $3.8^{+0.8}_{-0.7}$ & $279^{+12}_{-12}$ & $-19^{+11}_{-11}$ & $97 \pm 2$ & $130 \pm 20$ & Random \\
    \bottomrule
  \end{tabular}%
 } }
\newcommand{\PrintTabLocalE}{
  \resizebox{\textwidth}{!}{%
  \begin{tabular}{ccccccccc}
    \toprule
    $E_{\text{th}}$ (\unit{\EeV}) & $\Psi$ ($^\circ$) & TS & $S$ & $\ell$ ($^\circ$) & $b$ ($^\circ$) & $\mu_{E, \text{exp}}$ (\unit{\EeV}) & $\hat{\mu}_{E, \text{obs}}$ (\unit{\EeV}) & Probable source \\
    \midrule
    $8$ & $23^{+1}_{-3}$ & $23^{+7}_{-7}$ & $4.7^{+0.8}_{-0.8}$ & $309^{+8}_{-7}$ & $17^{+7}_{-7}$ & $14.83 \pm 0.04$ & $15.6 \pm 0.1$ & Cen~A \\
    $10$ & $24^{+1}_{-2}$ & $23^{+7}_{-7}$ & $4.7^{+0.8}_{-0.8}$ & $309^{+6}_{-6}$ & $17^{+6}_{-6}$ & $17.73 \pm 0.05$ & $18.7 \pm 0.2$ & Cen~A \\
    $13$ & $25^{+1}_{-3}$ & $22^{+7}_{-7}$ & $4.6^{+0.8}_{-0.8}$ & $309^{+6}_{-6}$ & $17^{+5}_{-5}$ & $21.98 \pm 0.08$ & $23.2 \pm 0.2$ & Cen~A \\
    $16$ & $21^{+1}_{-1}$ & $19^{+6}_{-6}$ & $4.3^{+0.7}_{-0.7}$ & $309^{+9}_{-8}$ & $16^{+8}_{-7}$ & $26.1 \pm 0.1$ & $27.9 \pm 0.3$ & Cen~A \\
    $20$ & $20^{+1}_{-1}$ & $16^{+5}_{-5}$ & $3.9^{+0.6}_{-0.6}$ & $309^{+10}_{-9}$ & $17^{+9}_{-8}$ & $31.2 \pm 0.1$ & $33.4 \pm 0.4$ & Cen~A \\
    $25$ & $20^{+1}_{-1}$ & $16^{+5}_{-5}$ & $3.9^{+0.6}_{-0.6}$ & $309^{+10}_{-10}$ & $17^{+9}_{-9}$ & $37.2 \pm 0.2$ & $40.2 \pm 0.5$ & Cen~A \\
    $32$ & $20^{+1}_{-1}$ & $13^{+3}_{-4}$ & $3.5^{+0.5}_{-0.5}$ & $309^{+10}_{-10}$ & $17^{+10}_{-10}$ & $44.7 \pm 0.3$ & $48.3 \pm 0.7$ & Cen~A \\
    $40$ & $20^{+1}_{-1}$ & $11^{+3}_{-3}$ & $3.3^{+0.5}_{-0.5}$ & $130^{+12}_{-12}$ & $-32^{+10}_{-10}$ & $52.9 \pm 0.4$ & $62 \pm 3$ & PPSC \\
    $50$ & $20^{+1}_{-1}$ & $11^{+3}_{-3}$ & $3.2^{+0.4}_{-0.5}$ & $272^{+11}_{-11}$ & $-13^{+11}_{-11}$ & $63.9 \pm 0.7$ & $73 \pm 2$ & Random \\
    $63$ & $20^{+1}_{-1}$ & $11^{+3}_{-3}$ & $3.2^{+0.4}_{-0.4}$ & $324^{+13}_{-13}$ & $-30^{+11}_{-11}$ & $79 \pm 1$ & $96 \pm 5$ & Random \\
    $79$ & $20^{+2}_{-1}$ & $11^{+3}_{-4}$ & $3.2^{+0.5}_{-0.5}$ & $339^{+12}_{-11}$ & $-14^{+12}_{-11}$ & $98 \pm 2$ & $130 \pm 10$ & Random \\
    \bottomrule
  \end{tabular}%
 } }
\section{Introduction}

Resolving the astrophysical sources of ultrahigh-energy cosmic rays (UHECRs) remains one of the most intricate open problems in contemporary astroparticle physics.
These particles, which reach energies well beyond $\qty{1}{\EeV} \equiv \qty{e18}{\eV}$, offer a unique probe into the Universe's most extreme environments where non-thermal particles are produced.
However, the search for their origins is severely complicated by the complex propagation of cosmic rays from their acceleration sites to their detection on Earth. As charged particles, UHECRs are deflected by the magnetic fields that permeate Galactic and extragalactic environments~\cite[e.g.,][]{open_questions,Unger2024}. 
The combination of largely unconstrained Galactic and extragalactic magnetic fields, along with the current difficulty in determining the cosmic-ray composition, challenges the prospects of UHECR astronomy~\cite{gmf_study,deOliveira_2022_egmf_agn,alvesbatista_uhecr_astronomy}.
Moreover, interactions with background photons~\cite{Greisen1966,Zatsepin1966,lang_revisiting} induce various energy losses, effectively establishing horizons beyond which the most energetic particles cannot reach Earth. While these horizons are highly composition-dependent, current observations impose a requirement for local sources within distances of $\lesssim\qty{100}{\Mpc}$~\cite{lang_revisiting}. 
Nevertheless, these estimates assume a standard scenario where the nuclear charge does not exceed that of iron, and the horizons can be considerably larger if UHECRs with energies around $\sim\!\qty{100}{\EeV}$ consist of heavier elements, as recently proposed~\cite{UH-UHECR,farrar_bns_2024}.

Despite these difficulties, the continuous operation of giant hybrid-detector arrays, such as the Pierre Auger~\cite{auger2015obs} and Telescope Array (TA)~\cite{TA2012} observatories, has yielded significant progress in obtaining more precise information about UHECRs. In particular, the study of UHECR arrival directions is a well-established approach that has provided valuable insights into their origins. On large angular scales, the discovery of a dipolar anisotropy by the Pierre Auger Collaboration~\cite{Aab2017,Auger2024} provides compelling evidence for an extragalactic origin of UHECRs with energies above $8\, \unit{\EeV}$.

At intermediate angular scales, the Auger Collaboration found evidence of correlations between UHECRs and starburst galaxies above $38\, \unit{\EeV}$~\cite{Auger2022}. 
Furthermore, for particles with energies
above $\sim\!\qty{40}{\EeV}$, a prominent excess has been observed in the direction
of the radio galaxy Centaurus A~\cite{Auger2022} which extends also to lower and higher energies~\cite{Auger2025sgp}.
When constraining the analysis to the supergalactic plane~\cite{Auger2025sgp}, the highest local significances of $5.2\,\sigma$ and $5.1\,\sigma$ are reached for $\Eth = \qty{20}{\EeV}$ and $\Eth = \qty{40}{\EeV}$, respectively, corresponding to post-trial significances of about $3\,\sigma$. When performing a fixed-target search in the Cen~A direction, the post-trial significance reaches $4.0\,\sigma$ for $\Eth \approx \qty{40}{\EeV}$~\cite{Geraldina2023ICRC}.

In the Northern Hemisphere, the TA Collaboration has similarly reported localized, intermediate-scale excesses at the highest energies. 
These include a hotspot in the direction of the Ursa Major constellation (known as the \textit{TA hotspot}~\cite{TA2014hotspot}) with a $5.1\sigma$ pre-trial significance \cite{TA2023hotspot} and another hotspot near the Perseus--Pisces supercluster (\textit{PPSC hotspot}~\cite{TA2023hotspot}) with a $3.8\sigma$ local significance. Although these intermediate-scale excesses represent a promising step toward source identification, they still lack the statistical significance required for a definitive discovery signal, having a local chance probability corresponding to $3.2\,\sigma$. Moreover, a robust correlation between the highest-energy events ($\gtrsim\!100\, \unit{\EeV}$) and potential source catalogs cannot yet be established~\cite{Bianciotto2025,TA2023Amaterasu,TA2024Heav,Bourriche_2026}.

Traditional studies of UHECR arrival directions rely mainly on identifying anisotropies in the particle flux across different angular scales and energy ranges. The emergence of these local particle excesses in different energy bands has prompted a deeper investigation into the directional energy distribution of UHECRs, inspiring new studies that combine energy spectrum and arrival-direction data. Recently, the Auger Collaboration performed a comprehensive search for spectral variations as a function of declination~\cite{Auger2025EnergySpectrum}. The study ultimately found no overall, statistically significant declination dependence in the energy spectrum across the Pierre Auger Observatory field of view. However, a mild spectral difference remains possible within the specific declination band containing the \textit{Centaurus region}. Specifically, a detailed study~\cite{Auger2025sgp} found that the excess of particles in this region exhibits a harder spectrum ($\gamma = 2.6 \pm 0.4$, above $20\, \unit{\EeV}$) compared to the overall energy spectrum within the Pierre Auger Observatory field of view ($\gamma = 2.99\pm 0.03$ for energies between $ 13\pm 1\, \unit{\EeV}$ and $48\pm 2\, \unit{\EeV}$, using slightly different reconstruction procedures~\cite{Auger2025EnergySpectrum}).
Interestingly, the TA Collaboration also found that the \textit{PPSC} and \textit{TA} hotspots exhibit a harder spectrum in the northern sky~\cite{TA2021Dec,TA2024Dec}; however, an independent analysis by the Pierre Auger Collaboration did not confirm the existence of a particle excess in that region~\cite{Auger2025sgp,Geraldina2023ICRC}.

Overall, these observations suggest that the \textit{Centaurus} and \textit{TA/PPSC} regions are characterized by distinct energy spectra. 
The discrepancies and features revealed by directional studies of the energy spectrum highlight the necessity of a method capable of unifying these recent results in a comprehensive and robust framework, potentially revealing other regions of spectral variations currently hidden across the sky. 

In this work, we introduce a new methodology to bridge this gap. 
We propose two novel metrics, the accumulated energy $U$ and the local average energy per particle $\mu_E$, which can be seamlessly integrated into analyses based on well-established methods for searching for flux anisotropies. These quantities fully account for the energy distribution of UHECRs in anisotropy studies across the sky and are easily applicable to both large- and small-scale analyses.

In \cref{sec:analytical_framework}, we compare the detection potential of anisotropies using three distinct metrics: particle flux ($\Phi_N$), energy flux ($\Phi_U$), and an estimator for the average energy ($\hat\mu_E$).
Because the accumulated energy $U$  inherently combines the flux overdensity and the presence of higher-energy particles, it amplifies the detection potential in regions where a particle excess is accompanied by a harder energy spectrum. To completely decouple spectral information from flux normalization, we demonstrate that $\hat\mu_E$ serves as a powerful estimator for tracing directionally dependent spectral features and is independent of flux intensity.
This indicates that $\hat\mu_E$ provides truly complementary information to the traditional arrival-direction approach, meaning their combination represents a powerful tool for uncovering potential signatures of UHECR sources.
Also, we show that, unlike the $N$- and $U$-metric, $\hat\mu_E$ is an intensive variable independent of the directional exposure for a continuous-operation observatory in the full-efficiency regime.
Moreover, unlike standard spectral analysis, this framework is model-independent, requiring no parametric fitting of the energy spectrum or prior assumptions regarding the functional form of the sources.

In \cref{sec:monte_carlo_simulation}, we apply the estimators to a simulated dataset and compare excesses in $N$, $U$, and $\hat\mu_E$ by generalizing the traditional Li--Ma~\cite{LiMa1983} significance formula.
For demonstration purposes, we assemble an exploratory model based on the parameters of Cen~A and \textit{PPSC} excesses. These regions are chosen since they are both in the field of view of the Auger Observatory and have an angular distance of $\sim\!170^\circ$, effectively preventing cross-contamination between the regions in this test scenario. The purpose is not to fully reproduce the sky seen by the Pierre Auger Observatory. 
For this reason, the results should not be interpreted as quantitative predictions for the UHECR sky, but rather as a demonstration of the analysis tool within a simple model.

By integrating over declination bands, we show in \cref{sec:integral_analysis} that the existence of local regions (intermediate-scale anisotropies) with a hard spectrum is fully compatible with global regions (large scales) without exhibiting variations in the spectral indices. 
This demonstrates that our formulation provides results consistent with those previously reported by the Pierre Auger Collaboration.
In \cref{sec:perspectives}, we discuss potential implications and possible astrophysical causes for regions with differing spectra. 
Finally, a summary of our conclusions is provided in \cref{sec:conclusion}.

\section{Methodology}
\label{sec:analytical_framework}

As previously noted, traditional methods employed in studies of arrival directions are generally based on searching for anisotropies in the particle flux ($\Phi_N$). In this section, we introduce and discuss the properties of the quantities proposed in our method: the accumulated energy ($U$) and the average energy ($\mu_E$). Consider a ground-based observatory in continuous operation with a fully efficient detector for particles arriving with local zenith angles $\theta\leq \theta_\tmax$ and energy above $E_{\rm min}$. In practice, this scenario applies directly to the surface detector array spaced \qty{1500}{\meter} apart (SD-1500) from the Pierre Auger Observatory for $E\geq \qty{4}{\EeV}$~\cite{auger2015obs}. Under these conditions, the observed number of particles per solid angle in a given right ascension and declination $(\alpha, \delta)$ is determined by the convolution of the cosmic-ray intensity $J(E;\alpha,\delta)$ at Earth with the Observatory exposure $\omega(E;\delta)$~\cite{sommers2001cosmic}, via
\begin{equation}
    \dv{N}{\Omega}\/(\alpha,\delta)  = \int_{\Eth}^{E_\tmax} J(E;\alpha,\delta) \omega(E; \delta) \dd{E},
    \label{eq:particle_flux}
\end{equation}where $\Eth$ is a given energy threshold of interest.
Since we are considering the full-efficiency regime, the exposure is independent of the energy, $\omega(E; \delta) \approx \omega(\delta)$, and can be safely factored out of the integral, giving
\begin{equation}
    \dv{N}{\Omega}\/(\alpha,\delta) = \Phi_N(\alpha,\delta) \omega(\delta),
    \label{eq:particle_integral_flux}
\end{equation}
where $\Phi_N(\alpha,\delta) = \int_{\Eth}^{E_\tmax} J(E;\alpha,\delta) \dd{E}$ is the integral particle flux.

To account for variations in the energy spectrum across the sky, it is advantageous to evaluate the accumulated energy $U\equiv\sum_i E_i$ from a given direction or region. 
Analogous to the particle flux, the accumulated energy  expected per solid angle is expressed as
\begin{equation}
    \dv{U}{\Omega}\/(\alpha,\delta) = \int_{\Eth}^{E_\tmax} E J(E;\alpha,\delta) \omega(\delta) \dd{E},
    \label{eq:energy_flux}
\end{equation}
giving
\begin{equation}
   \dv{U}{\Omega}\/(\alpha,\delta) = \Phi_U(\alpha,\delta) \omega(\delta),
   \label{eq:energy_flux_diff}
\end{equation}
where $\Phi_U(\alpha,\delta) = \int_{\Eth}^{E_\tmax} E J(E;\alpha,\delta) \dd{E}$ is the energy flux under the same assumptions for $\dv*{N}{\Omega}$.

The accumulated energy $U$ provides a significant advantage for anisotropy studies by naturally amplifying the statistical signal of the highest-energy particles. Because it inherently weights each event by its energy, $U$ becomes highly sensitive to the high-energy tail of the spectrum, potentially enhancing the significance of sources with hard spectra. However, it remains inherently correlated with the particle flux. For two regions with identical energy spectra, the region with a localized particle excess will naturally exhibit a higher total energy flux. While this correlation prevents a sky map of $U$ alone from isolating intrinsic spectral shapes, its sensitivity makes it a potential complementary tool.

To decouple spectral information from flux normalization and isolate intrinsic differences in the energy spectra from a given direction, we introduce a functional of the energy spectrum, defined as
\begin{equation}
    \mu_E(\alpha,\delta) \equiv \dv{U/d\Omega}{N /d \Omega} \/(\alpha,\delta) = \frac{ \int_{\Eth}^{E_\tmax}E J(E;\alpha,\delta) \omega(\delta) \dd{E}}{ \int_{\Eth}^{E_\tmax} J(E;\alpha,\delta) \omega(\delta) \dd{E}}
    \label{eq:expected_energy_def}
\end{equation}
 and interpreted as the average energy per particle in the direction $(\alpha, \delta)$.

 Note that, while $U$ potentially amplifies $N$-overdensities, $\mu_E$ acts as an intensive variable that depends solely on the functional form of the energy spectrum, independent of the flux intensity. 
Consequently, $\mu_E$ provides truly complementary information to maps of particle flux. 
Furthermore, because it is defined as a local ratio, $\mu_E$ remains independent of the directional exposure within the regime of full detection efficiency.

To evaluate $\mu_E$ in cosmic-ray data within a small solid angle $\Delta \Omega$, we use an estimator
\begin{equation}
    \hat{\mu}_E \equiv \frac{U/\Delta \Omega}{N/\Delta \Omega} = \frac{1}{N} \sum_{i=1}^{N} E_i \equiv \expval{E},
    \label{eq:energy_per_particle_def}
\end{equation}
where $\expval{\cdot}$ is the sample mean, with $N$ and $U$ being obtained in a solid angle of $\Delta \Omega  \approx 0.9^\circ \times 0.9^\circ \sim 10^{-4} \, \unit{\sr}$ compatible with the Auger Observatory resolution~\cite{auger2015obs}. For the next sections, it is also useful to define the true standard deviation, calculated via
\begin{equation}
    \sigma^2(E) = \Expval[E^2] - \Expval[E]^2, \text{ where } \Expval[E^2] \equiv \frac{\int_{\Eth}^{E_\tmax}E^2 J(E;\alpha,\delta) \dd{E}}{\int_{\Eth}^{E_\tmax} J(E;\alpha,\delta) \dd{E}},
    \label{eq:expected_energy_std_def}
\end{equation}
and its estimator
\begin{equation}
  \hat{\sigma}^2(E) = \frac{Q}{N} - \left(\frac{U}{N}\right)^2, \text{ where } Q \equiv \sum_i E_i^2.
  \label{eq:expected_energy_estimator_std_def}
\end{equation}

\section{Generalized anisotropies in the UHECR sky}
\label{sec:monte_carlo_simulation}

To evaluate the predictive power of $\mu_E$, we construct an exploratory model of the UHECR sky using Monte-Carlo simulations. 
We select parameters to model arrival-direction excesses and the energy spectrum based on results published by the Pierre Auger and Telescope Array collaborations.
To mimic observational conditions of a realistic experiment, we modulate the simulated particle flux by the directional exposure of the Pierre Auger Observatory, calculated following ref.~\cite{sommers2001cosmic}. 
Specifically, we adopt a latitude of $35.2^\circ \, \text{S}$, corresponding to the geographic location of the Pierre Auger Observatory~\cite{auger2015obs}, and restrict to air showers arriving at local zenith angles below $\theta_\tmax = 80^\circ$. Our goal is not to reproduce the observed UHECR sky precisely, but rather to explore the potential of the proposed method in a controlled and physically reasonable benchmark scenario.

\begin{figure}
    \centering
    \includegraphics[height=6cm]{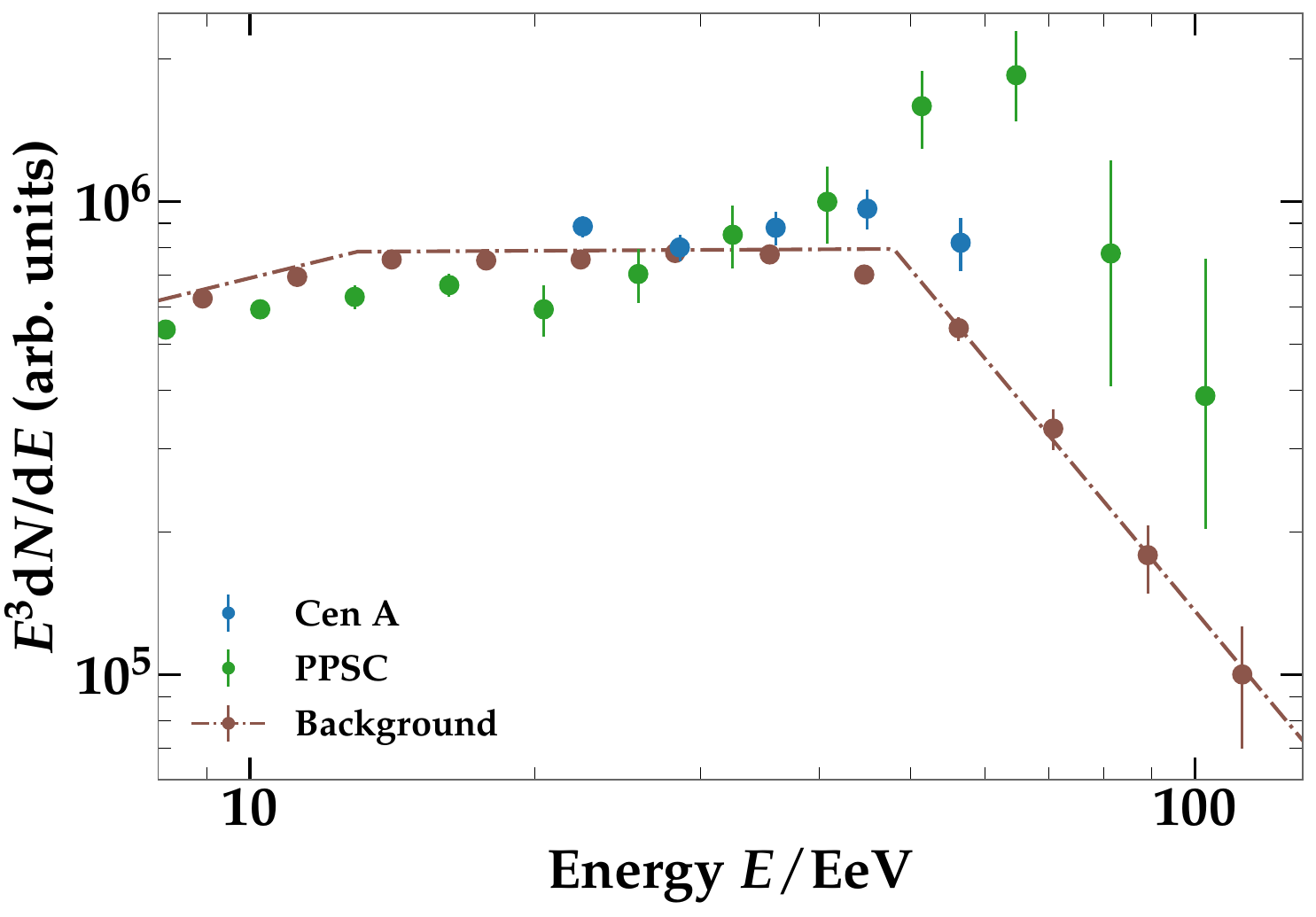}
    \caption{Energy spectra for the background (brown), PPSC (green), and Cen~A (blue), scaled by $E^3$ and normalized per unit solid angle. The background component follows the energy spectrum measured by the Pierre Auger Collaboration in their full field of view~\cite{Auger2025EnergySpectrum}. The Cen~A spectrum corresponds to the region within the Centaurus excess ($\Psi \approx 27^\circ$) observed by the Pierre Auger Collaboration~\cite{Auger2025sgp}. The spectrum assumed for the PPSC region is derived from the Northern sky excess ($\Psi \approx 20^\circ$) observed by the Telescope Array Collaboration~\cite{TA2024Dec}.}
    \label{fig:energy_spectrum}
\end{figure}

The model consists of three distinct components: an isotropic background, a localized particle excess above $20\, \unit{\EeV}$ within the \textit{Centaurus region}, and a region characterized by a harder energy spectrum in the \textit{PPSC hotspot} without an accompanying increase in the number of particles injected.
The minimum energy considered is set to $E_\tmin = \qty{8}{\EeV}$. The energy spectrum of each component is presented in \cref{fig:energy_spectrum}, and a 
detailed description is provided below.

The background component consists of particles isotropically distributed across the sky, whose energy spectrum follows the broken power-law reported by the Pierre Auger Collaboration over their overall field of view, given by~\cite{Auger2025EnergySpectrum}
\begin{equation}
J(E) \propto \begin{cases}
    E^{-2.5}, & \qty{8}{\EeV} \leq E < \qty{13}{\EeV},\\
    E^{-3.0}, & \qty{13}{\EeV} \leq E < \qty{48}{\EeV},\\
    E^{-5.4}, & E \geq \qty{48}{\EeV}.
\end{cases}
\label{eq:Auger_spectrum}
\end{equation}
For the purposes of this work, we neglect the dipolar modulation of the flux normalization (see ref.~\cite{Auger2025EnergySpectrum,Auger2020Spectrum}). 
Note that, to a first-order approximation, this modulation does not affect sky maps of $\hat\mu_E$. This holds within an energy interval in which the dipole amplitude and direction vary only slightly and follows from considering top-hat smoothing angles significantly smaller than the characteristic angular scale of the dipolar anisotropy, together with the intrinsic independence of $\hat{\mu}_E$ from the overall flux normalization.

The \textit{Centaurus region} is modeled as the combination of the background and a source-like component, resulting in a localized particle excess. 
The number of particles, energy spectrum, direction, and spatial extension of this component are adopted from the recent analysis by the Pierre Auger Collaboration~\cite[][table 3]{Auger2025sgp}. 
The source region is defined by a top-hat radius $\Psi \approx 27^\circ$ centered at $(\ell,b) \approx (309^\circ,17^\circ)$, containing $161$ excess particles injected with energies higher than \qty{20}{\EeV}. For comparison, 829 particles are attributed to the background within the same region. The total number of simulated particles in the field of view of the Observatory for $E\geq\qty{20}{\EeV}$, comprising both the signal excess and the background, is set to \num{8832} in accordance with ref.~\cite{Auger2025sgp}, and we assume that the particle contribution from Cen~A for $E<\qty{20}{\EeV}$ is negligible. The spectrum $\dv*{N}{E}$ is found by interpolating $N(\geq\!\Eth)$ in log-space from the values given in ref.~\cite{Auger2025sgp}; the results are similar to those obtained using a fitted power law.

In addition to the \textit{Centaurus region}, the \textit{PPSC} and \textit{TA hotspots} present compelling features to investigate, as the TA Collaboration reported them as local excesses of particles with Li--Ma significances of $\sim3.8 \sigma$ and $5.1\sigma$, respectively, both with a local chance probability equivalent to $\sim3.2\sigma$~\cite{TA2023hotspot} and a harder energy spectrum~\cite{TA2024Dec}. However, data from the Pierre Auger Observatory do not corroborate a particle excess or spectral hardening in these directions~\cite{Geraldina2023ICRC,Auger2025sgp}.

Motivated by this tension, we modeled a second excess in the UHECR sky based on the measurements of the northern hemisphere hotspots. We select the \textit{PPSC} excess, whose region spans an area slightly larger (relative to the total hotspot area) within the Pierre Auger Observatory field of view compared to the \textit{TA hotspot}. 
Note that, as their declinations are similar and the energy spectrum provided by TA corresponds to the sum of both regions, selecting the \textit{PPSC} excess over the \textit{TA hotspot} is expected to introduce only minor effects. Rather than attempting a full model of the excess, we simulate this test region directly from the energy spectrum of the \textit{TA/PPSC hotspots}, without separating it into source and background components. Consequently, we assume a harder energy spectrum but without an accompanying overall increase in particle flux at the injection energy ($\qty{8}{\EeV}$). This choice produces relative event excesses and corresponding deficits as a function of energy, even when the overall integrated flux matches the isotropic background. 
Our goal is not to reproduce the northern hemisphere hotspots, since the energy spectrum of each region is not available, and there is no agreement between TA and Auger measurements. We use the data and parameters from ref.~\cite{TA2023hotspot}, assuming a top-hat radius of $\Psi \approx 20^\circ$ centered on the reported excess direction $(\ell,b) \approx (127^\circ,-27^\circ)$. The hotspot spectrum $\dv*{N}{E}$ is found by interpolating $N(\geq\!\Eth)$ in log-space from the values given in ref.~\cite{TA2024Dec}; the results are similar to those obtained using the TA broken-power-law spectrum. The energies are converted to the Auger Observatory scale according to ref.~\cite{TA2024Dec}. 
This choice allows us to evaluate the sensitivity of the $\hat\mu_E$ estimator to hard-spectrum regions that lack a significant flux enhancement. 

\subsection{Analytical estimations}
\label{sec:analytical_estimations}

\begin{figure}
    \centering
    \includegraphics[width=1\linewidth]{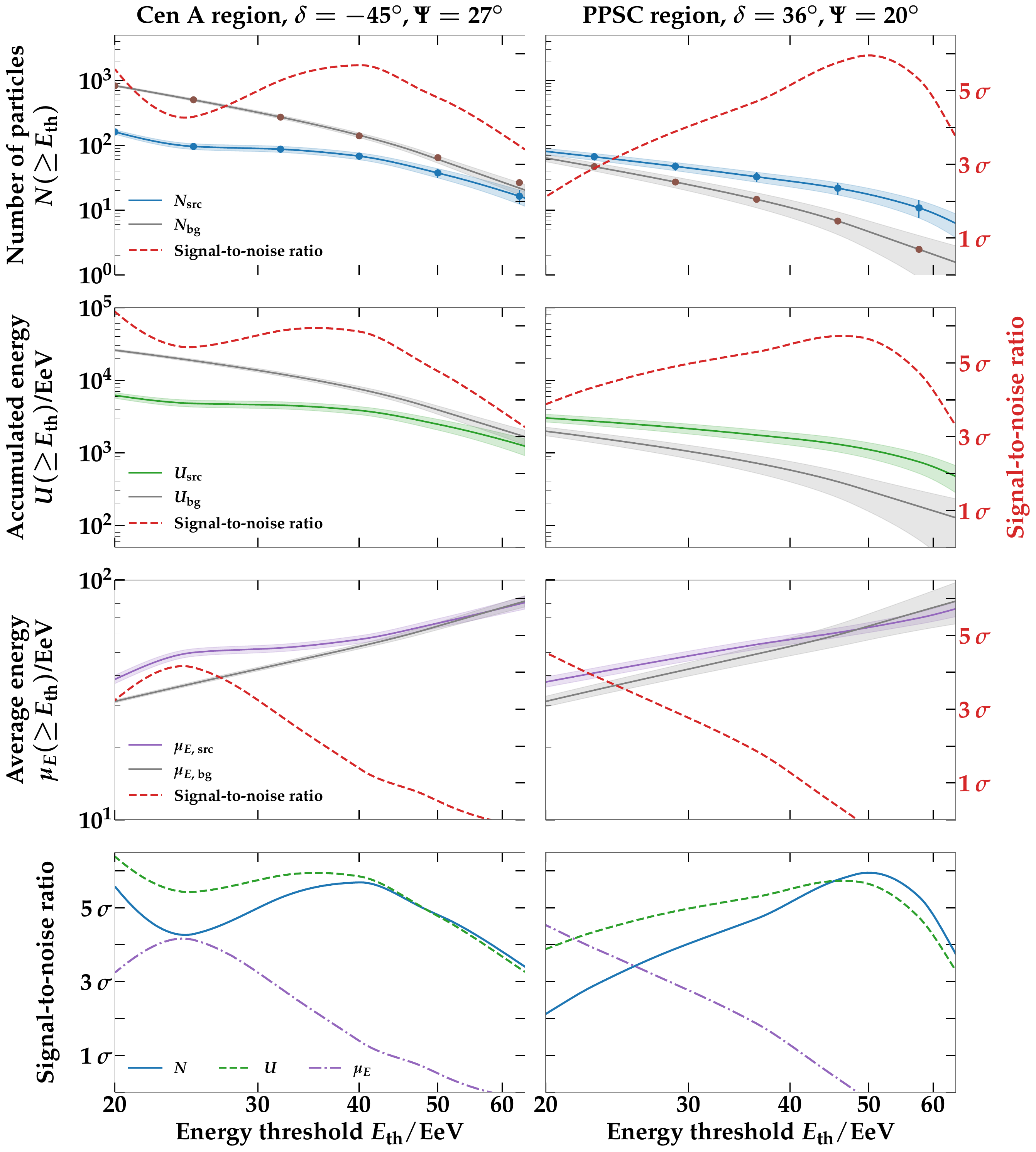}
    \caption{Analytical expectations for the $N(\geq\!\Eth)$ (upper), $U(\geq\!\Eth)$ (middle), and $\mu_E(\geq\!\Eth)$ (lower) in the Cen~A (left) and PPSC (right) regions originating from the background and from the sources. The secondary right axis (red) shows the signal-to-noise ratio approximation, also organized in the lower panel. The background for PPSC is defined as the number of particles expected for the PPSC region (direction and top-hat) if the energy spectrum follows \cref{eq:Auger_spectrum}. The results using the generalized Li--Ma significance are shown in \cref{fig:sig_vs_Eth}.}
    \label{fig:source_and_bg_vs_Eth}
\end{figure}

To get insight into the differences among $N$, $U$, and $\mu_E$, it is useful to perform an analytical study. Applying \cref{eq:particle_flux,eq:particle_integral_flux,eq:energy_flux,eq:expected_energy_def,eq:energy_per_particle_def,eq:expected_energy_std_def,eq:expected_energy_estimator_std_def,eq:energy_flux_diff} directly to the energy spectrum of each component (background, Cen~A, and PPSC) and integrating the directional exposure in the PPSC and Cen~A top-hat regions, we evaluated the distributions of $N$, $U$, and $\mu_E$ as a function of the threshold energy $\Eth$. 
The background spectrum for PPSC refers to the expected number of particles assuming \cref{eq:Auger_spectrum} in the PPSC top-hat region; a detailed explanation about the analytical integration is presented in \cref{app:analytical_integration}. The results are shown in \cref{fig:source_and_bg_vs_Eth}, where the left and right panels present the expectations for Cen~A and PPSC, respectively. 

To provide a first-order estimation of the source significance in each metric, the signal-to-noise ratio for each quantity ($s_N$, $s_U$, and $s_E$, from top to bottom) is also displayed on the right axes. The approximations are performed by explicitly integrating each signal and noise. It recovers the true significance in the weak-source $N_\source \ll N_\background$ limit; a detailed explanation is provided in \cref{app:significance_approximation} and discussed in further sections. Finally, the bottom panels compare $s$ directly across all three quantities. The curves are restricted to $\Eth \ge \qty{20}{\EeV}$ to prevent unphysical signal-to-noise artifacts caused by extrapolating below the source emission threshold. Additionally, the broader PPSC uncertainty bands result from its harder energy spectrum, which increases intrinsic variance, combined with a lower particle count.

Considering the variable $N$ (\cref{fig:source_and_bg_vs_Eth}, top panels), the spectral features are clearly reflected in the predicted values of $s_N$. 
For the Cen~A component (left panels), the structure of the significance curve is consistent with the findings of the Pierre Auger Collaboration reported in ref.~\cite{Auger2025sgp}, exhibiting a double-peaked profile with local significances exceeding $\!5\,\sigma$ at $\sim\!20$ and $\sim\!\qty{40}{\EeV}$.
Note that the search for excesses in the full field of view is normally done by the Auger Collaboration by defining a pre-trial $p$-value using a binomial distribution, which is afterward penalized to compute the post-trial significance; a Li--Ma significance is normally used with the best-fit parameters only for visualization. In the case of ref.~\cite{Auger2025sgp}, the search was confined to the supergalactic plane, obtaining local Li--Ma significances of $5.2\,\sigma$ for $E\geq \qty{20}{\EeV}$ and $5.1\,\sigma$ for $E\geq \qty{40}{\EeV}$, both at $(\ell,b) = (309^\circ,17^\circ)$, the Cen~A direction. The agreement between the signal-to-noise ratio found in our exploratory model and the results from ref.~\cite{Auger2025sgp} indicates that this region has been modeled consistently with the observed data. 

In the case of the PPSC region (right panels), the energy spectrum closely matches the background model for $E \lesssim 40\text{ EeV}$.
At $\Eth \sim 50\, \unit{\EeV}$, the excess reaches $s_N \sim 6\, \sigma$, substantially higher than the $2.4\sigma$ local Li--Ma observed in the Auger Observatory data~\cite{Auger2025sgp}. The high signal-to-noise obtained is attributed to the hard spectral index assumed for the PPSC region, which, coupled with the constrained total number of particles, inevitably produces a strong relative flux enhancement at high energies.
As shown in \cref{sec:N_anisotropies}, the local significance decreases to $\sim\!5\,\sigma$ once Poisson fluctuations in finite event samples are accounted for via the Li--Ma significance formula~\cite{LiMa1983}. Despite this discrepancy for higher energies, this exploratory model remains a suitable benchmark for exploring the discovery potential of the different metrics in the UHECR sky.

The results obtained for $U$ are very similar to those for $N$, both qualitatively and quantitatively. However, as shown in the bottom panels of \cref{fig:source_and_bg_vs_Eth}, the significance achieved for $U$ exceeds that for $N$ across both the Cen~A and PPSC regions. This enhancement occurs because $U$ is inherently more sensitive to the high-energy tail of the spectrum than $N$, thereby amplifying spectral differences at higher energies and consequently increasing the statistical significance (see the discussion in \cref{sec:U_anisotropies}).

In turn, the profile of $\mu_E$ differs significantly from those of $N$ and $U$. High values of $\mu_E$ indicate high ratios of $U$ to $N$. Unlike $s_N$ and $s_U$, $s_E$ is maximized at lower $\Eth$.
Inflections in $s_E$ correspond directly to energy thresholds where the spectral index departs from that of the background. This behavior is clearly illustrated by the anti-correlated peaks of $s_E$ relative to $s_N$ and $s_U$ for Cen~A, as well as the sharp drop in $s_E$ for PPSC.

\subsection[Monte-Carlo simulations]{Monte-Carlo simulations}

To perform an analysis in the full field of view of the Auger Observatory, we generate $\Nsample$ mock datasets by Monte-Carlo sampling the arrival-direction and energy distributions under the conditions described above. For each realization, we compute sky maps of $N$, $U$, and $\hat\mu_E$ according to \cref{eq:particle_integral_flux,eq:energy_flux,eq:expected_energy_def}, and their respective Li--Ma significances ($S$; see below). The simulated cosmic-ray arrival directions are pixelated using a \textsc{Healpix}\footnote{\url{https://healpix.sourceforge.io/}.} grid with \num{49152} pixels~\cite{healpix2005,healpy2019}, corresponding to an angular resolution of $\sim\!0.9^\circ$, comparable to the angular resolution of the Pierre Auger Observatory.

To evaluate the significance $S_N$, we employ the standard and widely used Li--Ma framework (eq.~(17) in ref.~\cite{LiMa1983}). This formulation is appropriate for determining the significance of an excess in the particle count $N$ observed within a top-hat search window of radius $\Psi$ centered on a given direction $(\alpha,\delta)$ with local exposure $\EE = \int \omega(\delta) \dd{\Omega}$. However, this method is suitable only for discrete counting data governed by Poisson statistics and cannot be applied directly to continuous observables such as $U$ and $\hat\mu_E$, which follow complex or \textit{a priori} unknown probability distributions.\footnote{Because our goal is to establish a model-independent framework, we do not assume a specific functional form for the energy spectrum, such as a pure power law or a power law with an exponential cutoff, when calculating the significances.} To overcome this limitation, we generalize the Li--Ma significance formula to account for both excesses and deficits in the integrated energy flux $U$ across the sky, as (for details, see \cref{app:li-ma})
\begin{equation}
	S_U \equiv \sqrt{\TS_U} =  \frac{U_\on - \alpha U_\off}{\sqrt{\alpha (Q_\on + Q_\off)}}, \quad \alpha \equiv \mathcal{E}_\on/\mathcal{E}_\off,
    \label{eq:LiMa_U}
\end{equation}
where the ``on'' (``off'') label denotes particles within the on-source (off-source) region, and $Q$ is defined in \cref{eq:expected_energy_estimator_std_def}. Analogously, we extend this formulation to evaluate the significance in the average energy per particle, $\hat{\mu}_E$, expressed as
\begin{equation}
    S_E \equiv \sqrt{\TS_E} = \frac{\hat\mu_\text{$E$,on} - \hat\mu_\text{$E$,off}}{\sqrt{\frac{\hat\sigma^2(E)}{N_\on}\left(1+\frac{1}{N_\on}\right) + \frac{\hat\sigma^2(E)}{N_\off} \left(1+\frac{1}{N_\off}\right) }}, \quad \hat\sigma^2(E) =  \frac{Q_\off}{N_\off} - \left(\frac{U_\off}{N_\off}\right)^2.
    \label{eq:LiMa_E}
\end{equation}

The generalized Li--Ma significances are computed for different energy thresholds $\Eth$ and top-hat search radii $\Psi$. We evaluate energy thresholds spanning from $8$ to $\sim\!\qty{80}{\EeV}$, using equally spaced logarithmic steps of $\Delta \log_{10}(E/{\rm EeV}) = 0.1$, rounded to the nearest EeV. For each $\Eth$, a maximum-likelihood search is performed over top-hat radii $\Psi = 20^\circ,21^\circ,\ldots,30^\circ$; extending the search window up to $\Psi = 45^\circ$ was also evaluated but yielded no statistically significant gains.

\begin{figure}
    \centering
    \includegraphics[width=1\linewidth]{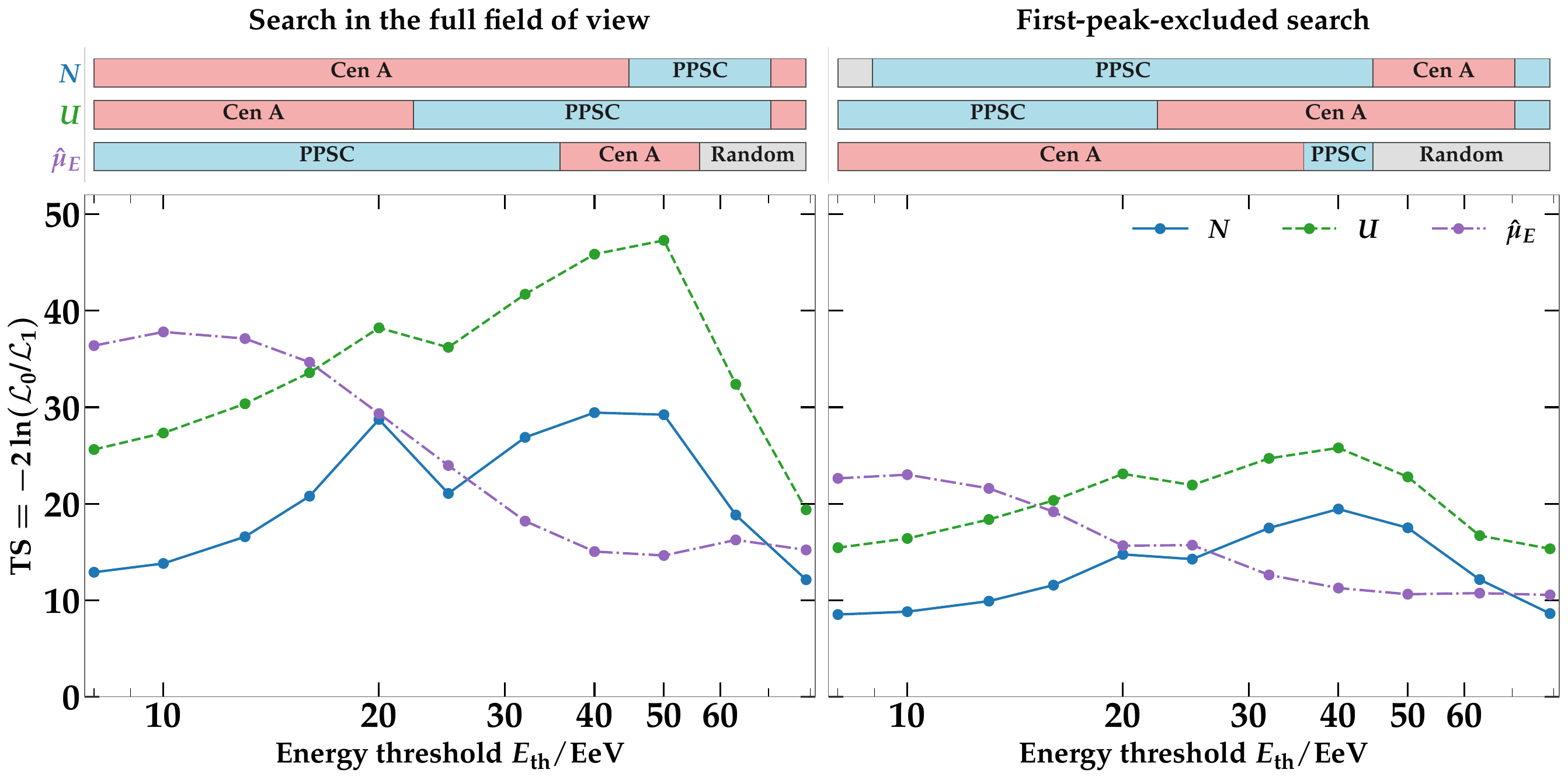}
    \caption{Test statistic TS for $N$, $U$, and $\hat\mu_E$ averaged over $\Nsample$ simulations. The left panels show the results obtained when searching the full field of view of the Auger Observatory. The right panels show those obtained when excluding the regions found with the full-sky search. The uncertainties are not displayed for the sake of readability. The upper panels show whether the signal direction is within the top-hat region of a source (red for Cen~A and blue for PPSC) or if it is in a random direction (gray).}
    \label{fig:sig_vs_Eth}
\end{figure}

The dependence of the test statistic (TS) on $\Eth$ for $N$, $U$, and $\hat\mu_E$ is presented in \cref{fig:sig_vs_Eth}; detailed tables of results are provided in \cref{app:extra_tables}. 
The analysis was conducted both over the full field of view (referred to as full-sky) and by masking the region of maximum excess identified in the full-sky search. 
In the full-sky analysis, the variation of TS with $\Eth$ is similar to the analytical signal-to-noise ratio predictions (\cref{fig:source_and_bg_vs_Eth}). 
However, notable differences arise due to transitions between the dominant sources, which can be seen in \cref{tab:local_N,tab:local_U,tab:local_E,tab:global_N,tab:global_U,tab:global_E} and in the upper panels of \cref{fig:sig_vs_Eth}. We attribute a signal to a given source (Cen~A or PPSC) when its reconstructed direction lies within the respective source's top-hat region; when the direction falls outside both regions, the \textit{random} label is used.

The peak excess remains predominantly within the top-hat regions of the sources.
A few exceptions occur for $\hat\mu_E$ at high $\Eth$, where sample sizes are limited at higher energies, and for $N$ at $\Eth = 10\, \unit{\EeV}$ (when the primary peak is excluded), attributed to the spectral similarity between the background and PPSC. Overall, the inferred values of $\Psi$ reported in \cref{tab:local_N,tab:local_U,tab:local_E,tab:global_N,tab:global_U,tab:global_E} are consistent with the input parameters ($20^\circ$ for PPSC and $27^\circ$ for Cen~A).

\begin{table}
  \centering
  \caption{Results for the maximum likelihood analysis averaged over $\Nsample$ simulations. For comparison, Cen~A is defined at $(\ell,b) \approx (309^\circ,17^\circ)$ within a top-hat angle $\Psi = 27^\circ$, and PPSC at $(\ell,b) \approx (127^\circ,-27^\circ)$ with $\Psi = 20^\circ$.}
  \vspace{3mm}
  \label{tab:results_TS}
  \resizebox{1\textwidth}{!}{%
 \PrintTabResultsTS
  }
\end{table}

The results obtained from the maximization process for $S_N$, $S_U$, and $S_E$ are summarized in \cref{tab:results_TS}. We report the mean values across all realizations for the search radius $\Psi$, test statistic (TS), generalized Li--Ma significance ($S$), reconstructed excess direction $(\ell, b)$, the $N$, $U$, and $\hat\mu_E$ expected for the null hypothesis (calculated using the off-source region), and the  $N$, $U$, and $\hat\mu_E$ found in the on-source region. Uncertainties denote the 68\% confidence level across all simulations; the dispersion between different simulations is presented in \cref{sec:dispersion_simulations}. For $N$ and $U$, both the global and local maxima identified in \cref{fig:sig_vs_Eth} are listed.

\begin{figure}
    \centering
    \includegraphics[width=\textwidth]{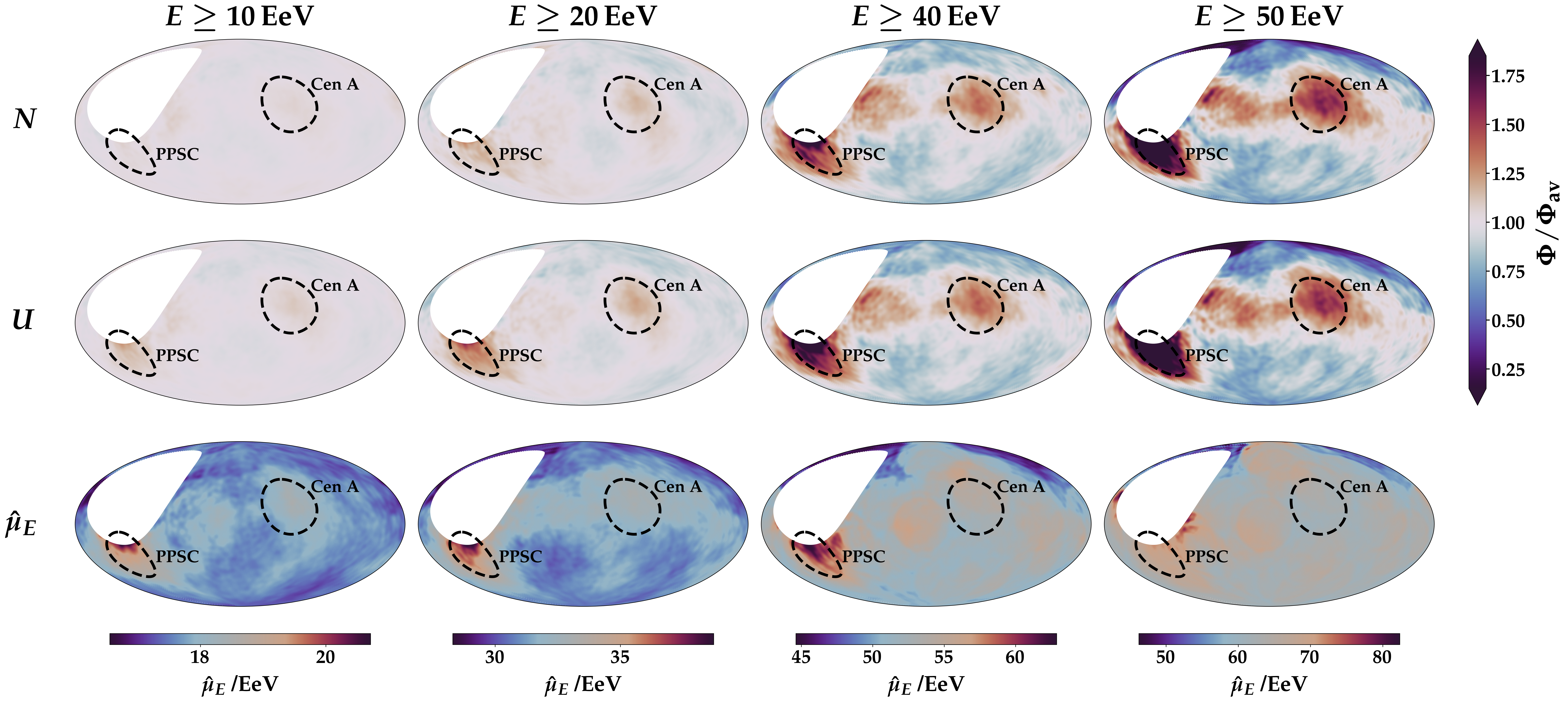}
     \caption{Sky map in Galactic coordinates using a Hammer projection showing the $N$ (upper) and $U$ (center) flux divided by the average value and $\hat\mu_E$ (lower) in $\unit{\EeV}$ in a representative mock dataset for a top-hat radius $\Psi = 27^\circ$.}
    \label{fig:flux_map}
\end{figure}

\begin{figure}
    \centering
    \includegraphics[width=0.95\textwidth]{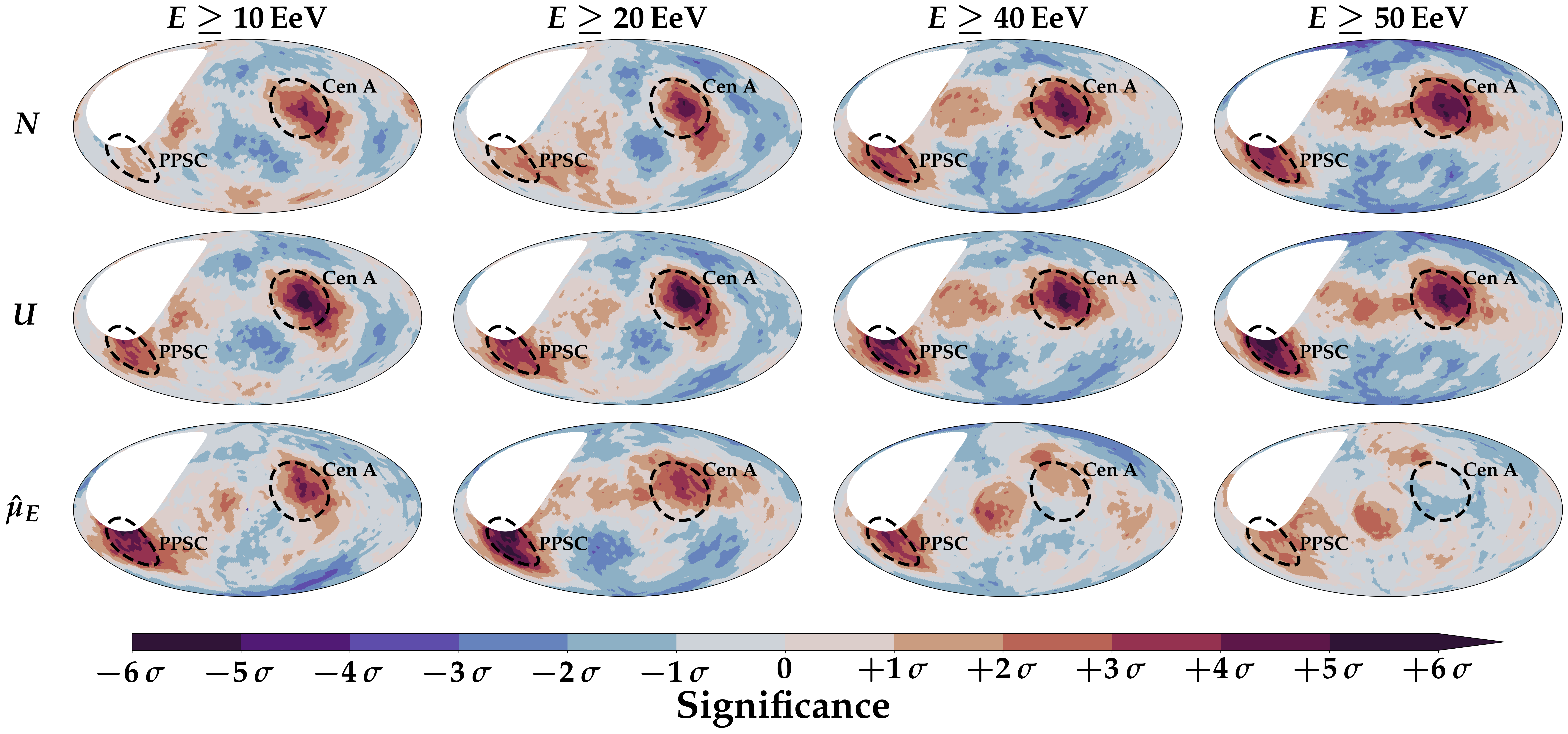}
     \caption{Sky map in Galactic coordinates using a Hammer projection showing the generalized Li--Ma significance for the $N$ (upper), $U$ (center), and $\hat\mu_E$ (lower) metrics in a representative mock dataset for a top-hat radius $\Psi = 27^\circ$.}
    \label{fig:li_ma_map}
\end{figure}

Sky maps of the integrated particle and energy flux for a representative realization adopting the optimized parameters are presented in \cref{fig:flux_map} (top and center panels), alongside the corresponding sky maps of $\hat\mu_E$ (bottom panels). The four columns display results for energy thresholds $\Eth = 10, 20, 40,$ and $50\, \unit{\EeV}$. A top-hat smoothing angle of $27^\circ$ is applied. The respective generalized Li--Ma significances for $N$, $U$, and $\hat{\mu}_E$ are shown in \cref{fig:li_ma_map}. In the following, we discuss the results obtained for each metric.

\subsubsection[Anisotropies in the particle flux (\texorpdfstring{$N$}{N}-anisotropies)]{\texorpdfstring{Anisotropies in the particle flux ($\bm{N}$}{N}-anisotropies)}
\label{sec:N_anisotropies}

As the energy threshold increases, excesses in particle flux originating from the Cen~A and PPSC regions become clearly distinguishable in the $N$-flux sky maps (\cref{fig:flux_map}).
The Cen~A region largely dominates the significance sky map when evaluated using the $N$-metric. As detailed in \cref{tab:global_N}, all local significance peaks up to $\Eth = 40\, \unit{\EeV}$ are centered on the Cen~A direction, $(\ell, b) \approx (309^\circ, 17^\circ)$. For $\Eth=\qty{50}{\EeV}$ and $\Eth=\qty{63}{\EeV}$, however, the PPSC region dominates the all-sky search, yielding local significances of $5.3\,\sigma$ and $4.3\,\sigma$, respectively. This transition in excess dominance is also clearly captured in \cref{fig:sig_vs_Eth}, where the shift directly impacts the profile of TS as a function of  $\Eth$. Comparing these trends with \cref{fig:source_and_bg_vs_Eth}, we observe that the TS evolution follows the $s_N$ profile for Cen~A up to $\Eth \sim 40\, \unit{\EeV}$. However, whereas $s_N$ drops at $40\text{ EeV}$, TS forms a plateau driven by the emergence of PPSC as the dominant source.

In \cref{fig:li_ma_map}, we note the presence of two significant peaks in the all-sky search: the aforementioned $5.4\sigma$ peak at $\Eth =\qty{40}{\EeV}$ and a secondary $5.3\sigma$ peak for $\Eth =\qty{20}{\EeV}$, both located within the Cen~A region. This profile agrees with the expectations from the analytical calculations for Cen~A (\cref{fig:source_and_bg_vs_Eth}). 
Both the $\sim\!\qty{40}{\EeV}$ and the $\sim\!\qty{20}{\EeV}$ peaks are in accordance with the local significance found by the Auger Collaboration in ref.~\cite{Auger2025sgp}, as discussed in \cref{sec:analytical_estimations}. 
For the $20\, \unit{\EeV}$ threshold, $\num{828\pm30}$ particles are expected while $\num{995\pm 42}$ are observed. 
When masking the top-hat region of the primary peak, the remaining signal is dominated by the PPSC region, yielding significances of 3--$4\,\sigma$ for $\Eth = \qty{20}{\EeV}$ and 4--$5\,\sigma$ for $\Eth = 40\, \unit{\EeV}$, as detailed in \cref{tab:local_N}.
Overall, the search for anisotropies in the particle flux demonstrates the self-consistency of the exploratory model developed for this study, as well as the prominent role played by Cen~A when the $N$-metric is considered.

\subsubsection[Anisotropies in energy flux (\texorpdfstring{$U$}{U}-anisotropies)]{\texorpdfstring{Anisotropies in energy flux ($\bm{U}$}{U}-anisotropies)}
\label{sec:U_anisotropies}

Qualitatively, the distribution of the accumulated energy $U$ exhibits a pattern very similar to that observed for $N$, as shown in \cref{fig:flux_map}. This similarity arises because regions containing a larger number of particles naturally accumulate a higher total energy sum, $U \equiv \sum_i E_i$. However, $S_U$ is significantly higher than $S_N$. Assuming that the background dominates the total variance, the significance scales as\footnote{$\sigma^2(U_\background) = N_\background \Expval[E^2]_\background$; see details in \cref{app:li-ma,app:significance_approximation}.}
\begin{equation}
    S_U \approx \frac{\hat\mu_\text{$E$, source}}{E_{\text{rms, bg}}} S_N, \text{ where } E_\text{rms} \equiv \sqrt{\expval{E^2}},
\end{equation}
showing that, in this regime, the $U$-significance corresponds to the standard $N$-significance scaled by a spectral factor. For $\hat\mu_\text{$E$, source} > E_{\text{rms, bg}}$, $S_U$ enhances the significance, thereby enhancing the discovery potential of high-energy sources (see \cref{fig:sig_vs_Eth,fig:source_and_bg_vs_Eth}).

However, unlike the behavior observed for $N$, the two TS peaks obtained as a function of $\Eth$ point to different sources. For $\Eth = \qty{20}{\EeV}$, a $6.1\sigma$ peak is found in the Cen~A direction, where a total energy sum of $\qty{25900+-800}{\EeV}$ is expected and $\qty{32000+-1000}{\EeV}$ is observed; when penalized for the search in different regions of the sky using different top-hat angles and threshold energies, we have a post-trial significance of $3.2\,\sigma$. For $\Eth =  \qty{50}{\EeV}$, we find a $6.7\sigma$ peak (post-trial: $3.6\,\sigma$) in the PPSC direction, where we set up a harder spectrum; in this region, $\qty{330+-90}{\EeV}$ is expected and $\qty{1300+-300}{\EeV}$ is found. When excluding the first-peak region (see \cref{tab:local_U}), the Cen~A excess at $E\geq\qty{40}{\EeV}$ is recovered with a local significance of $5.0\,\sigma$.

\subsubsection[Anisotropies in average energy (\texorpdfstring{$\mu_E$}{mu_E}-anisotropies)]{Anisotropies in average energy (\texorpdfstring{$\bm{\mu_E}$}{E}-anisotropies)}

Similar to the behavior observed in \cref{sec:analytical_estimations}, the sky maps of $\hat\mu_E$ exhibit significant changes compared to the particle or energy flux maps (see \cref{fig:li_ma_map,fig:sig_vs_Eth}). 
The significance $S_E$ is maximized for lower energy thresholds around $\Eth \sim 10\, \unit{\EeV}$. When higher energies are considered ($\sim\!50\, \unit{\EeV}$), the significances are largely erased.
As reported in \cref{tab:global_E}, the significance is dominated by PPSC at lower energies ($\Eth<\qty{32}{\EeV}$), featuring a $6.1\sigma$ peak at $\Eth = \qty{10}{\EeV}$, corresponding to a post-trial significance of $3.5\,\sigma$ after penalization. In this energy range, an average energy of $\hat\mu_E=\qty{17.78+-0.05}{\EeV}$ is expected for the isotropic background spectrum (see \cref{eq:Auger_spectrum}); a value of $\hat\mu_E = \qty{21.5+-0.8}{\EeV}$ is found, as shown in the energy map displayed in \cref{fig:flux_map}. When excluding the first-peak region (see \cref{tab:local_E}), a $4.7\sigma$ peak is found in the Cen~A direction at $\Eth = \qty{10}{\EeV}$, where $\qty{17.73+-0.05}{\EeV}$ is expected and $\qty{18.7+-0.2}{\EeV}$ is found. As shown in \cref{fig:li_ma_map,fig:flux_map}, the $\hat\mu_E$-metric strongly highlights the PPSC region, although the Cen~A $4.7\sigma$ peak is also visible.

Note that, under a hypothesis of a weak source, the $\hat\mu_E$ signal behaves as\footnote{See the derivation in \cref{app:significance_approximation}.}
\begin{equation}
    S_E \approx \frac{\hat\mu_\text{$E$, source} - \hat\mu_\text{$E$, bg}}{\hat\sigma(E)} |S_N|.
    \label{eq:estimation_significance_E}
\end{equation}
Although $\mu_E$-excesses do not intrinsically depend on a $N$-excess, as shown by \cref{eq:energy_per_particle_def}, the signal in the generalized Li--Ma formula increases as the statistical uncertainty in the energy decreases, justifying the $|S_N|$ factor. If the energy spectra of the source and background are identical, we find $\hat\mu_\text{$E$, source} =\hat\mu_\text{$E$, bg}$, leading to $S_E =0$. Analogously, if the source exhibits a softer spectrum compared to the background, we find $S_E<0$, whereas for a harder spectrum, $S_E>0$.

\subsection{Impact of extreme-energy events and energy uncertainties}
\label{sec:bias_extreme}

\begin{figure}
    \centering
    \includegraphics[width=1\textwidth]{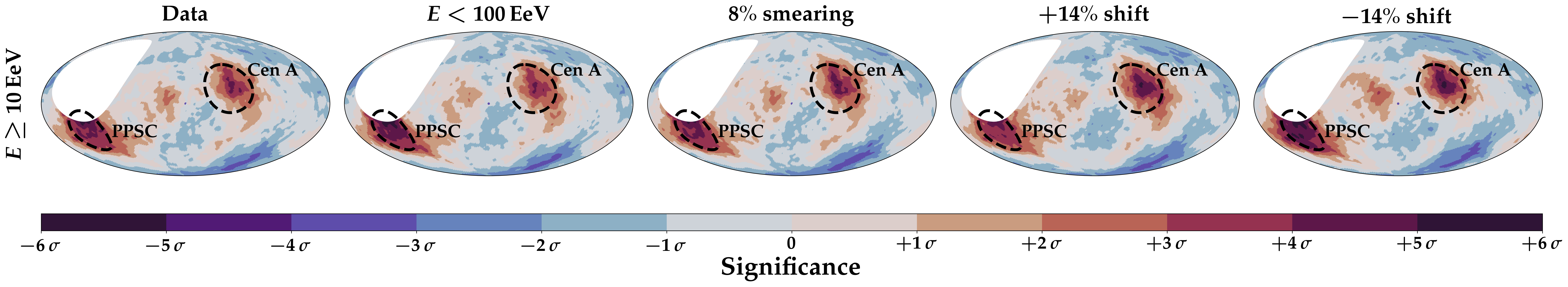}
    \caption{Sky map in Galactic coordinates using a Hammer projection showing the generalized Li--Ma significance in the $\hat\mu_E$-metric for a representative mock dataset for $E\geq\qty{10}{\EeV}$ using a top-hat radius $\Psi = 27^\circ$. The first panel shows the representative sky map of the mock dataset; the second shows the result when removing data with $E\geq \qty{100}{\EeV}$; the third shows the result when smearing the energy by 8\% using a Gaussian beam; the fourth (fifth) shows the result when shifting all the energies by $+14\%$ ($-14\%$).}
    \label{fig:E_smearing_and_extreme}
\end{figure}

We investigate whether extreme-energy events and energy uncertainties (resolution and systematics) could bias the estimator $\hat\mu_E$ when calculating the significance $S_E$. In \cref{fig:E_smearing_and_extreme} (first and second panels), we compare the generalized Li--Ma significance sky maps for the highest $\mu_E$-excess presented in \cref{fig:li_ma_map} with and without events above \qty{100}{\EeV}. Overall, the qualitative behavior of the sky maps remains consistent between both energy cuts.
Removing $E > \qty{100}{\EeV}$ events leads to only modest relative changes in the significance, increasing the global maximum significance for $E\geq \qty{10}{\EeV}$ by 7\%, without shifting the sky locations of the excesses. 
Because the signal structure and spatial localization remain highly consistent at the optimal lower thresholds, we conclude that the detection with $\hat{\mu}_E$ is robust and not artificially driven by a handful of extreme-energy events.

In the energy range considered in this paper, the systematic uncertainty on the absolute energy scale for the Pierre Auger Observatory \textsc{Phase I} data is $\sim\!14\%$~\cite{auger2015obs}, while the energy resolution is below $\sim\!8\%$~\cite{Auger2020Spectrum}. We evaluate whether the uncertainty in energy from the Auger Observatory could bias our technique. First, to account for energy resolution, we randomly smear the energies of the simulated particles using a Gaussian beam with an $8\%$ standard deviation. Separately, to assess the impact of the systematic uncertainty, we apply a $\pm14\%$ shift to the energies of all events in the mock dataset. The results, presented in \cref{fig:E_smearing_and_extreme} for a representative dataset, demonstrate that the technique is robust against both effects. Whether applying statistical smearing or systematic shifts, the spatial location of the highest signals remains stable, and the absolute value of the maximum significance fluctuates by only $(3\text{--}7)\%$ relative to the baseline data.

\section{Integral analysis}
\label{sec:integral_analysis}

\begin{figure}
    \centering
    \includegraphics[width=0.95\linewidth]{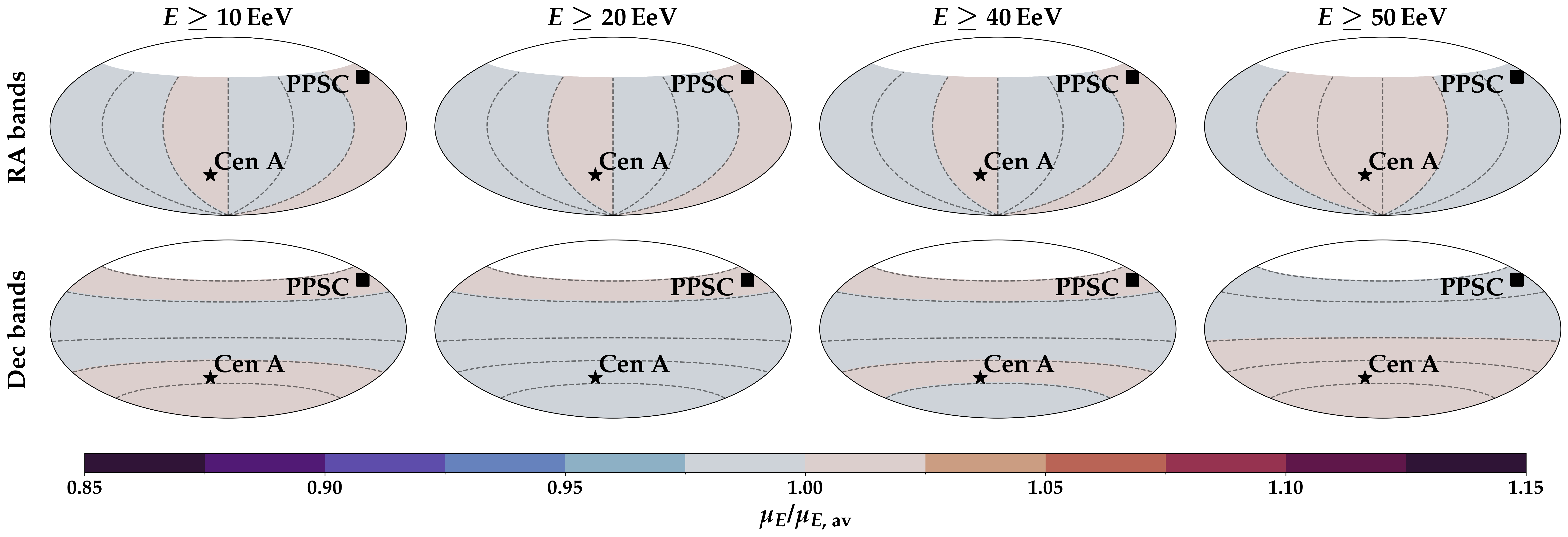}
    \caption{Sky map in equatorial coordinates using a Hammer projection showing average energy computed by integrating the particle and energy flux in right-ascension and declination bands over \num{10000} simulations; the values are normalized by dividing by the average value in the sky map. The coordinate system is defined with $\alpha=0^\circ$ on the right and increasing to the left.}
    \label{fig:integral_analysis}
\end{figure}

In the preceding sections, we demonstrated that the \textit{Centaurus} and \textit{PPSC} regions can manifest as intermediate-scale anisotropies in $\hat\mu_E$. This behavior is intrinsically connected to localized variations in the energy spectrum, which could potentially be identified in observational UHECR data.

Recently, the Pierre Auger Collaboration~\cite{Auger2025EnergySpectrum} performed a search for declination dependence in the energy spectrum by dividing the sky into five declination bands: $[-90^\circ,-51^\circ]$, $[-51^\circ, -29^\circ]$, $[-29^\circ, -8.0^\circ]$, $[-8.0^\circ, 24.8^\circ]$, $[24.8^\circ, 44.8^\circ]$. The specific bin spanning declination $24.8^\circ$ to $44.8^\circ$ was selected as it contains exclusively inclined events (local zenith angle $\theta$ between $60^\circ$ and $80^\circ$), while the remaining four bins were defined to have similar exposures. Their results indicated that the spectra across different bands are statistically compatible, exhibiting only a mild spectral deviation in the direction of the \textit{Centaurus region}.

To assess whether the presence of intermediate-scale $\mu_E$-anisotropies is consistent with the absence of significant spectral differences in declination bands, we evaluated localized spectral variations within our exploratory model using the $\hat\mu_E$ functional across the declination bands defined in ref.~\cite{Auger2025EnergySpectrum}. The analysis was performed by integrating the $N$ and $U$ fluxes within specified declination bands, subsequently computing $\hat\mu_E$ via \cref{eq:energy_per_particle_def}; the same analysis is also performed in right-ascension bands. The results are summarized in \cref{tab:RA_integration,tab:Dec_integration} and displayed in \cref{fig:integral_analysis} for four energy ranges. 

The integrated average energies across all declination bands remain statistically compatible within $1\,\sigma$.
When integrating over right-ascension bands, small deviations of up to $2\,\sigma$ emerge in the intervals encompassing Cen~A $(\alpha,\delta) \approx (201^\circ,-45^\circ)$ and PPSC $(\alpha,\delta) \approx (17^\circ,36^\circ)$, particularly for $\Eth = \qty{10}{\EeV}$; however, such deviations are not unexpected given the number of angular bins tested.
Note that, while the local analysis in \cref{sec:monte_carlo_simulation} yields $\delta \hat\mu_E \sim 2\text{--}4\, \unit{\EeV}$ for Cen~A and PPSC, these differences diminish to $\lesssim 0.2\, \unit{\EeV}$ when integrated over declination bands. 
This confirms that intermediate-scale $\mu_E$-anisotropies become strongly diluted when averaged over large spatial bands, rendering their existence fully compatible with the absence of significant band-integrated spectral variations.

\begin{table}
    \centering
    \caption{Average energy $\hat\mu_E$ after integration over right ascension in declination bands.}
    \PrintTabResultsRA
    \begin{flushleft}
        \footnotesize
        *Includes Cen~A declination $\delta \approx -45^\circ$.\\
        **Includes PPSC declination $\delta \approx 36^\circ$.
    \end{flushleft}
    \label{tab:RA_integration}
\end{table}

\begin{table}
    \centering
    \caption{Average energy $\hat\mu_E$ after integration over declination in right-ascension bands.}
    \PrintTabResultsDec
    \begin{flushleft}
        \footnotesize
        *Includes Cen~A right ascension $\alpha \approx 201^\circ$.\\
        **Includes PPSC right ascension $\alpha \approx 17^\circ$.
    \end{flushleft}
    \label{tab:Dec_integration}
\end{table}

\section{Astrophysical implications and perspectives for source detection} \label{sec:perspectives}

As a functional of the energy spectrum, $\mu_E$ represents an intrinsic spectral property. In this section, we investigate how the injection spectrum and the propagation impact $\mu_E$ from the perspective of searching for UHECR sources. 
In a scenario where such sources exhibit diverse injection parameters, variations in $\mu_E$ across the sky can naturally be attributed to the dominant contribution of specific sources within a given region. Moreover, even a scenario featuring identical sources will yield spatial variations in the energy spectrum due to modulations induced by extragalactic propagation. Consequently, the resulting $\mu_E$-map depends not only on the intrinsic spectral features of a source, but also on its distance, injected composition, temporal injection profile, and intervening magnetic fields. Furthermore, multiple sources may contribute to the observed events in a given sky region; in such cases, the relative contribution among these sources plays a crucial role in determining the overall energy spectrum and, consequently, $\mu_E$.

\begin{figure}
    \centering
    \includegraphics[height=5.5cm]{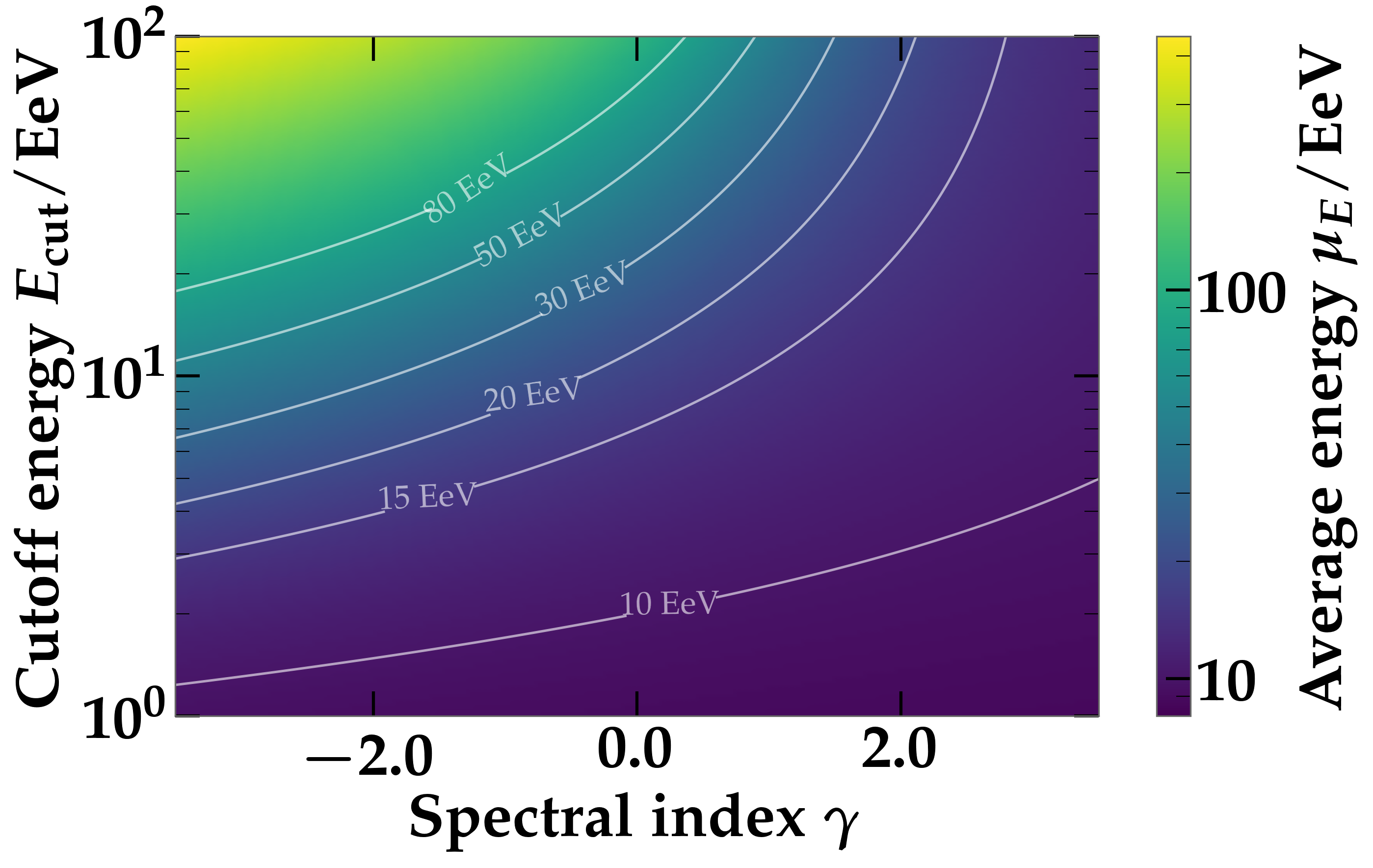} 
    \caption{Average energy per particle as a function of the spectral index $\gamma$ and cutoff energy $E_\cut$ for a flux $\propto \ee^{-E/E_\cut}E^{-\gamma}$, assuming $E\geq \qty{8}{\EeV}$. The contour lines represent different $\mu_E$ in EeV.}
    \label{fig:energy_vs_gamma}
\end{figure}

\begin{figure}
    \centering
    \includegraphics[width=\linewidth]{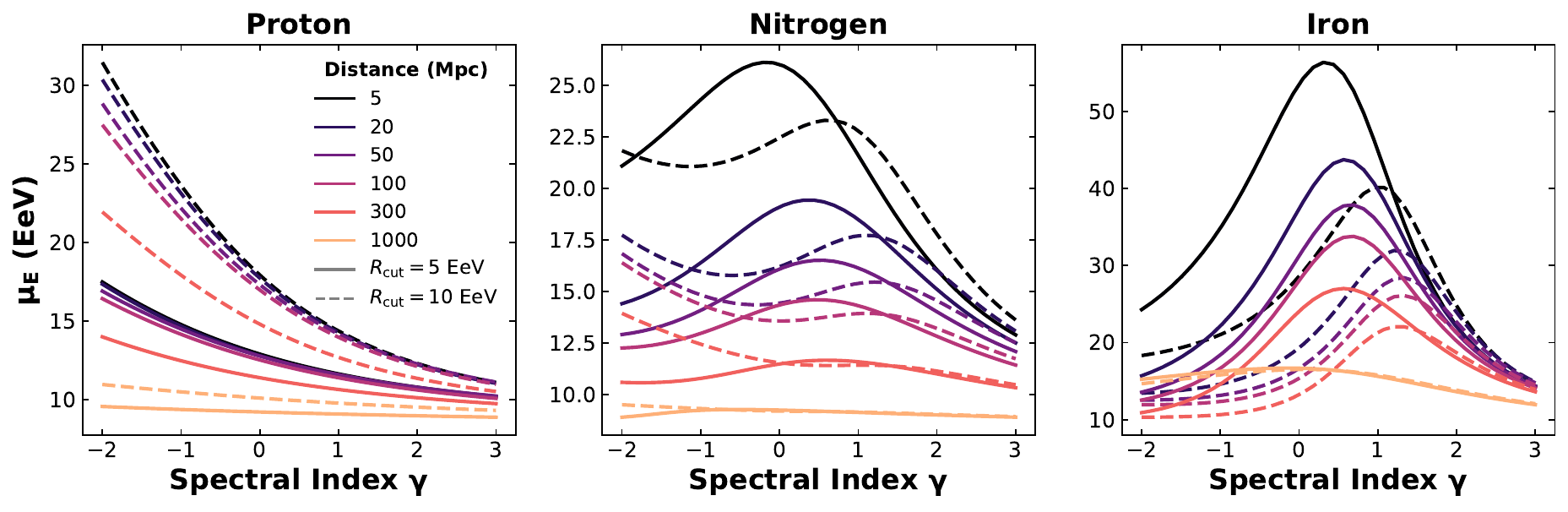}
    \caption{Average energy per particle as a function of the spectral index obtained for three injected charges: protons (left), nitrogen (center), and iron (right). Several injection distances (colors) and two $R_{\rm cut}$ (continuous and dashed lines) are shown. All nuclear fragments arriving at the observer are shown.}
   \label{fig:Eavg_prop}
\end{figure}

To evaluate the effects of energy attenuation and mass composition, we performed one-dimensional simulations of UHECR propagation using the \textsc{CRPropa3.2}\footnote{\url{https://crpropa.github.io/CRPropa3/}.} framework~\cite{CRPropa2022}. 
The simulations incorporate all relevant attenuation processes, including photo-pion and pair productions, photodisintegration, nuclear decay, and adiabatic losses due to cosmic expansion. 
The background photon fields included are the cosmic microwave background (CMB) and the extragalactic background light (EBL), where the model provided by Gilmore \textit{et al.}~\cite{Gilmore2012} was employed.
Photodisintegration cross-sections were computed using TALYS~\cite{Koning2005,Koning2012}.
Events were sampled uniformly in logarithmic energy space and subsequently reweighted such that the source injection spectrum follows a power law with an exponential cutoff, $E^{-\gamma} \exp(-E / E_{\rm cut})$.

In \cref{fig:energy_vs_gamma}, we illustrate the behavior of $\mu_E$ at the source as a function of the intrinsic injection parameters $\gamma$ and $E_{\rm cut}$. As expected, $\mu_E$ increases for harder energy spectra. The average energy also rises with increasing cutoff energy, which effectively extends the upper limit of the integration range. The influence of $E_{\rm cut}$ becomes less pronounced for larger values of $\gamma$, where the energy distribution is predominantly concentrated at lower energies.

However, attenuation processes during UHECR propagation modify $\mu_E$ considerably.
\Cref{fig:Eavg_prop} shows the evolution of $\mu_E$ as a function of the spectral index for point sources at various distances ($D$). Results considering three representative injection species (protons, nitrogen, and iron) are evaluated for two cutoff rigidities, $R_{\rm cut} = 5$ and $10\, \unit{\EeV}$, assuming $E_{\rm cut} = Z R_{\rm cut}$, where $Z$ is the primary nuclear charge. 
For positive spectral indices, $\mu_E$ increases as the spectrum becomes harder and for higher values of $R_{\rm cut}$, as expected from the injection spectrum. Particularly, in this regime, $\mu_E$ scales with the primary charge. However, for nuclei, the harder spectrum ($\gamma \lesssim 0$) considerably increases the number of primaries subjected to intense photodisintegration. The rapid production of lower-energy secondaries shifts the mean energy downward and compensates for the decrease in $\gamma$, producing a peak in $\mu_E$. The properties of the source spectrum become less pronounced for longer propagation distances. In the limit of very distant sources ($\sim\!1000\, \unit{\Mpc}$), this propagation-induced cutoff completely dominates the intrinsic source properties (initial $E_{\rm cut}$ and $\gamma$).

In addition to the attenuation effect considered, the temporal injection profile of UHECRs also induces significant changes in the observed energy spectrum and, consequently, in $\mu_E$. Several studies~\cite[e.g.,][]{flickering_jets,Eichmann_2023,Mollerach_cenA,deOliveira_history,temporalinvarianceillusion} have investigated the effects of a finite source lifetime on the energy spectrum. Due to time delays induced by propagation in magnetic fields, particles of higher rigidity diffuse faster than those with lower rigidity. This implies that the signal from sources whose activity has recently begun tends to exhibit a harder spectrum than that of long-lived sources. Conversely, the opposite behavior is expected for sources that have already deactivated (see, e.g., \cite{deOliveira_history}). Both temporal signatures should be imprinted in $\mu_E$. This effect may be less pronounced if the rigidity is approximately constant with energy or in the presence of very weak magnetic fields.

\section{Conclusion}
\label{sec:conclusion}

In this work, we propose a general framework to identify sky anisotropies associated with localized variations in the energy spectrum, which potentially point to the presence of UHECR sources. Due to attenuation effects during UHECR propagation, spectral anisotropies are intimately linked to local source distributions. The spatial distribution of event arrival directions has long been the primary tool for identifying localized sources~\cite[e.g.,][]{Auger2022,Aab2017,kascade_grande2019,icecube2016,eastop2009}, leveraging propagation horizons produced by energy loss processes to constrain source distances at the highest energies. However, propagation also introduces modifications across the entire energy spectrum by driving energy losses and generating secondary particles. Additional factors, such as temporal emission profiles coupled with Galactic and extragalactic magnetic fields, further shape the observed spectra.

Localized excesses of UHECR particles ($N$) become increasingly pronounced at higher energies. This trend is supported by several observations, including localized excesses in the Cen~A region ($\gtrsim\!\qty{40}{\EeV}$) and the Perseus-Pisces supercluster ($\gtrsim\!\qty{30}{\EeV}$), and correlations with starburst galaxies for energies exceeding $\qty{38}{\EeV}$~\cite{Auger2018,Auger2022,Geraldina2023ICRC}. 
The emergence of such excesses at higher energies suggests the presence of nearby sources revealed at the highest energies, where magnetic deflections are less significant. The determination of the energy spectra of the \textit{Centaurus} and \textit{PPSC} regions indicates that they exhibit not only an excess of particles, but also distinct spectral shapes. Taken together, these features point to a potentially emergent and complementary source signature.

To characterize a given sky region, we propose coupling the search for event-number anisotropies ($N$-anisotropies) with energy-flux anisotropies ($U$-anisotropies) and average-energy anisotropies ($\mu_E$-anisotropies). 
While the $U$-metric enhances the discovery potential by amplifying the signal of high-energy tails, the $\hat\mu_E$-metric isolates intrinsic spectral features, without requiring model-dependent spectral fits. 
Spatial variations in these energy-weighted metrics might reflect proximity to UHECR sources, propagation attenuation, or deflections in Galactic and extragalactic magnetic fields. 

To evaluate the capability of $N$, $U$, and $\mu_E$ to characterize localized anisotropies, we constructed a data-inspired model of the UHECR sky. It combines the primary localized excesses observed in arrival directions (\textit{Centaurus} and \textit{PPSC} regions) with their respective energy spectra. To assess the statistical significance of each quantity relative to an isotropic sky, we derived a generalization of the standard Li--Ma significance formula extended to $U$ and $\mu_E$. Our method has the potential to identify regions of spectral variation that would otherwise remain hidden in standard particle-flux maps.

The \textit{Centaurus region} hosts the most prominent local excess observed by the Pierre Auger Collaboration~\cite{Auger2022}. The exploratory model investigated in section~\ref{sec:monte_carlo_simulation} indicates that Cen~A should dominate $N$-anisotropies, but also be detectable in $U$- and $\mu_E$-anisotropies. For $\Eth \sim 20\, \unit{\EeV}$, the significance $S_U$ can surpass that of $S_N$, whereas slightly lower values are expected for $S_E$.

The \textit{PPSC} and \textit{TA} hotspots have been discussed in the debate regarding the compatibility of the energy spectra measured by the Pierre Auger and TA observatories in the Southern and Northern skies, respectively. The results obtained by the TA Collaboration~\cite{TA2024Dec} indicate that the spectrum in the \textit{PPSC} and \textit{TA} hotspots differs from the background spectrum. Note that, while applying an energy-dependent correction is likely to reconcile the spectra in the common-declination band~\cite{AugerTASpectrum2024}, it cannot explain spectral differences inferred from the data of each collaboration independently. However, it is also important to highlight that the detection of the hotspots by the Telescope Array Collaboration has a chance probability equivalent to a $\sim3\sigma$ significance level.
Our exploratory model indicates that a region with no overall excess of particles (no integrated $N$-anisotropies) but presenting a harder spectrum -- as in the simplified model applied to the \textit{PPSC} region -- can still manifest $U$- and $\mu_E$-anisotropies. Although the observed significance in real data may not reach the high values predicted by our model ($S_N \sim 4.4\,\sigma$, $S_U \sim 6.7 \,\sigma$, and $S_E \sim 6.1\, \sigma$ pre-trial and $S_N\sim 1.2\,\sigma,S_U\sim 3.6\,\sigma,$ and $S_E\sim 3.5\,\sigma$ post-trial), the enhancement of $S_E$ relative to $S_N$ indicates that such a feature could be identified as a hard-spectrum region in realistic searches to establish the \textit{PPSC region} as a potential astrophysical source.

The combination of $N$-, $U$-, and $\mu_E$-anisotropies can provide further insights into the UHECR sky beyond discoveries based solely on $N$-anisotropies.
While $N$ and $U$ exhibit similar spatial excess patterns, $U$ yields systematically higher statistical significances for hard-spectrum sources. Furthermore, distinct spectral signatures emerge for $\mu_E$, where prominent excesses appear at lower energies.
The methodology proposed here has the potential to:
\begin{itemize}
    \item Enhance the statistical significance of localized excesses with hard energy spectra by leveraging $U$-maps, which naturally amplify high-energy signatures;

    \item Identify small- and intermediate-scale regions with hard energy spectra that become explicit in $\mu_E$-maps;
    \item Provide exposure-independent spectral metrics in the full-efficiency regime, as the local ratio $\hat\mu_E = U/N$ cancels out the directional exposure function $\omega(\delta)$;
    \item Search for anisotropies in the energy distribution without requiring model-dependent fits to the energy spectrum in each region of interest;
    \item Help to distinguish between true astrophysical origins and detector energy-scale calibrations;
    \item Uncover localized spectral variations that are otherwise obscured when integrating over declination or right-ascension bands.
\end{itemize}

Two key consistency tests were conducted in this study. First, in \cref{sec:bias_extreme}, we verified that, although $\mu_E$ depends on energy, it is not driven by the presence of a few cosmic rays with extreme energies and remains robust against differences in the energy calibration or statistical variations due to the energy resolution.

Second, in \cref{sec:integral_analysis} we demonstrated that the pronounced $>\!5\sigma$ pre-trial $\mu_E$ variations associated with localized excesses are smoothed into minor perturbations when the spectrum is integrated over right-ascension or declination bands. 
Consequently, localized regions of spectral variance, such as \textit{Cen~A} and \textit{PPSC}, remain fully compatible with the energy spectrum measurements in integrated declination bands reported by the Pierre Auger Collaboration~\cite{Auger2025EnergySpectrum}. 
This raises the possibility that additional, hitherto unknown, regions of spectral variation exist, forming hidden structures across the UHECR sky. Promising candidates include particle excess regions potentially associated with Fornax~A~\cite{Auger2018,matthews_fornax, deOliveira_2022_egmf_agn, martins2026uhecrdoubletsconditionalassociation}; the TA \textit{hotspot} and \textit{coldspot}~\cite{Abbasi_2018_ta_hotspot, kim2026recentfindingstelescopearray}; and the directions of starburst galaxies, as reported in a joint publication by the Pierre Auger and TA Collaborations~\cite{Urena2025}, where a $4.2\sigma$ post-trial significance was found. Starburst galaxies are considered compelling candidates for UHECR acceleration due to their potentially enhanced transient activity. Alternatively, the magnetic fields of starburst environments may echo cosmic-ray signals emitted by other sources, such as Cen~A~\cite{echoes_radiogal,echos_council_giants}. Combining $N$-, $U$-, and $\mu_E$-anisotropies offers a promising metric for probing these regions and distinguishing intrinsic acceleration from an echoed signal (e.g., by comparing their spectral profiles with that of Cen~A). Future studies will be essential to evaluate the feasibility of this strategy.

Such precision studies are made possible by the unprecedented statistics achieved by the Pierre Auger and Telescope Array Observatories. Looking forward, incorporating mass composition estimates into this framework will enable rigidity-based anisotropy searches using the complementary pair ($R, \mu_R$), including possible assemblies of rigidity sky maps.

The framework proposed here can be naturally extended to:
\begin{itemize}
    \item Independently measure spectral differences across bands in different coordinate systems (e.g., Galactic or Supergalactic);
    \item Search for large-scale energy-space anisotropies, such as dipole and quadrupole moments in both $U$ and $\mu_E$. For instance, comparing the dipole and quadrupole moments of $N$, $U$, and $\mu_E$, or evaluating correlations of $\mu_E$ and $U$ with the Galactic and supergalactic planes, could efficiently decouple standard particle-flux variations from spectral signatures;    %
    \item Evaluate energy-weighted correlations with astrophysical source catalogs;
    \item Correlate $N$-, $U$-,  and $\mu_E$-anisotropies to disentangle intrinsic source activity from magnetic field echoes and potential transient emission signatures. For magnetic echoes, there is likely a connection between the direct trajectory of the signal and its delayed echoes. However, modulations driven by attenuation processes and diffusive propagation along extended trajectory path lengths make this comparison indirect, requiring detailed propagation modeling. For nearby recent transients, the observed spectrum can exhibit distinct spectral features due to energy-dependent time delays and the initial arrival of a ballistic flux front.
\end{itemize}

Ultimately, this framework represents an advancement in our ongoing efforts to uncover the origin of UHECRs. Rather than replacing well-established particle-count searches, these energy-weighted metrics provide a complementary approach to the existing analytical toolkit. Consequently, they can significantly enhance our ability to identify the specific astrophysical environments responsible for accelerating the most energetic particles in the Universe.

\section*{Acknowledgments}

We thank Carola Dobrigkeit, Rogerio de Menezes, Edivaldo Moura Santos, Geraldina Golup, Marta Bianciotto, Miguel Martins, Bruce Dawson, and Teresa Bister for fruitful discussions and comments. We acknowledge ``Centro Nacional de Processamento de Alto Desempenho em São Paulo (CENAPAD-SP)'' for the HPC resources. This study was financed, in part, by the São Paulo Research Foundation (FAPESP), Brasil. Process Number 2025/03325-5, 2021/01089-1, and 2019/10151-2.

\appendix
\crefalias{section}{appendix}
\makeatletter
\@addtoreset{table}{section}
\@addtoreset{figure}{section}
\makeatother
\renewcommand{\thetable}{\thesection.\arabic{table}}
\renewcommand{\thefigure}{\thesection.\arabic{figure}}

\section{Generalization of the Li--Ma formula}
\label{app:li-ma}

The original Li--Ma significance framework employs a likelihood-ratio test tailored to the discrete Poisson nature of particle counts $N$. Formulating an analogous likelihood ratio for the energy domain would require assuming specific parametric models for the source and background spectra (e.g., pure power laws). 
By the Central Limit Theorem, in the limit of a sufficiently large number of events, the distributions of the continuous energy variables, namely the accumulated energy $U$ and the average energy per particle $\hat\mu_E$, can be approximated by Gaussian distributions.

We can derive model-independent formulae for the significance of energy excesses using a standard likelihood-ratio test under this Gaussian approximation.
Let $X$ represent our generic continuous variable (either $U$ or $\hat\mu_E$). 
We define the likelihood function $\LL$ for measuring the actual values $X_\text{on}$ and $X_\text{off}$ in the source and background regions, respectively, given their true expected values $\mu_\text{on}$ and $\mu_\text{off}$, and their variances $ \sigma_\on^2 \equiv \sigma^2(X_\text{on})$ and $\sigma_\off^2 \equiv \sigma^2(X_\text{off})$, via
\begin{equation}
    \LL(X_\text{on}, X_\text{off} \mid \mu_\text{on}, \mu_\text{off}) = G(X_\text{on} \mid \mu_\text{on}, \sigma_\text{on}) \times G(X_\text{off} \mid \mu_\text{off}, \sigma_\text{off}),
\end{equation}
where $G(x \mid \mu, \sigma) \equiv \frac{1}{\sqrt{2\pi}\sigma} \exp\left[-\frac{(x-\mu)^2}{2\sigma^2}\right]$. Thus,
\begin{equation}
    \ln \LL(X_\text{on}, X_\text{off} \mid \mu_\text{on}, \mu_\text{off}) = -\frac{1}{2} \left[ \frac{(X_\on-\mu_\on)^2}{\sigma_\on^2} + \frac{(X_\off - \mu_\off)^2}{\sigma_\off^2} \right] + C,
\end{equation}
where $C$ is a constant. Under the alternative hypothesis ($\tH_1$), where a source is present, the likelihood is trivially maximized at $\hat{\mu}_\on = X_\on$ and $\hat{\mu}_\off = X_\off$, giving
\begin{equation}
    \ell_1 \equiv \ln \left[\sup_{\text{ $\mu_\on$, $\mu_\off$ under $\tH_1$}}\LL(X_\text{on}, X_\text{off} \mid \mu_\text{on}, \mu_\text{off})\right] = C,
\end{equation}
where sup notation refers to the supremum.

Under the null hypothesis ($\tH_0$), where there is no source, the expected value in the on region is strictly proportional to the expected value in the off region by some scaling factor $k$, while $\mu_\off$ is a probe for the background $\mu_\background$, giving
\begin{equation}
    \tH_0: \ \left\{\begin{aligned}
        & \mu_\off = \mu_\background, \\
        &  \mu_\on = k \mu_\off = k\mu_\background.
    \end{aligned}\right.
\end{equation}
Hence, the maximum likelihood can be found by 
\begin{equation}
    \pdv{\ln \LL_0}{\mu_\background} = \frac{X_\on - k\mu_\background}{\sigma_\on^2} k + \frac{X_\off - \mu_\background}{\sigma_\off^2} = 0,
\end{equation}
yielding a best-fit background estimator
\begin{equation}
    \hat{\mu}_\background = \frac{k \sigma^2_\off X_\on + \sigma^2_\on X_\off}{k^2 \sigma_\off^2 + \sigma_\on^2}.
\end{equation}
Therefore, the logarithm of the maximized likelihood function under $\tH_0$ is
\begin{equation}
    \ell_0 \equiv \ln \left[\sup_{\text{$\mu_\on$, $\mu_\off$ under $\tH_0$}}\LL(X_\text{on}, X_\text{off} \mid \mu_\text{on}, \mu_\text{off})\right] = - \frac{1}{2} \frac{(X_\on-kX_\off)^2}{\sigma_\on^2 + k^2 \sigma_\off^2} + C.
\end{equation}
If we define the signal as $X_S \equiv X_\on - k X_\off$ and the noise as $\sigma_S \equiv \sigma(X_S) = \sqrt{\sigma_\on^2 + k^2 \sigma_\off^2}$, we find a test statistic
\begin{equation}
    \TS \equiv -2 (\ell_0 - \ell_1) = X_S^2/\sigma_S^2.
\end{equation}
By Wilks's theorem~\cite{Wilks1938}, $\TS$ follows a $\chi^2$ distribution with one degree of freedom, giving us a significance
\begin{equation}
    S \equiv \sqrt{\TS} = \frac{X_S}{\sigma_S}.
\end{equation}
Hence, under our assumptions of Gaussian distributions, the significance in $U$ or $\hat\mu_E$ can be estimated by the signal-to-noise ratio. The validity of the Gaussian assumption is discussed in \cref{app:validity_gauss}.

\subsection[\texorpdfstring{$U$}{U}-metric]{\texorpdfstring{$\bm{U}$}{U}-metric}

We define $U_\on$ and $U_\off$ as the energy sums inside and outside a certain region
\begin{equation}
	U_\on \equiv \sum_{i=1}^{N_\on} E_i, \quad U_\off \equiv \sum_{i=1}^{N_\off} E_i.
\end{equation}
Also,
\begin{equation}
	U_\tot \equiv \sum_{i=1}^{N_\tot} E_i = U_\on + U_\off \text{ and } N_\tot \equiv N_\on + N_\off.
\end{equation}
The energy accumulated due to the source is
\begin{equation}
	U_S = U_\on - \alpha U_\off,
\end{equation}
where $\alpha \equiv \EE_\on/\EE_\off$, and $\EE_\on$ ($\EE_\off$) is the integrated exposure in the on (off) region.

Given a variance
\begin{equation}
	\sigma^2(U_S) = \sigma^2(U_\on) + \alpha^2 \sigma^2(U_\off),
\end{equation}
the signal-to-noise ratio is calculated by
\begin{equation}
	{S_U = \frac{U_\on - \alpha U_\off}{\sqrt{\sigma^2(U_\on) + \alpha^2 \sigma^2(U_\off)}}.}
\end{equation}
Since $U_\off \equiv \sum_{i=1}^{N_\off} E_i$ is a Compound Poisson process,\footnote{A compound Poisson process is the sum of independent and identically distributed random variables $E_i$, where the number of terms $N_\off$ in the sum is itself a Poisson-distributed random variable.} $\sigma^2(U_\off) = \Expval[N_\off] \Expval[E^2]$; hence, the estimator for the variance is
\begin{equation}
	\left\{ \begin{aligned}
	    & \hat{\sigma}^2(U_\off) = \sum_{i=1}^{N_\off} E_i^2 \equiv Q_\off,\\
        & 	\hat{\sigma}^2(U_\on) = \sum_{i=1}^{N_\on} E_i^2 \equiv Q_\on,
	\end{aligned}\right.
\end{equation}
where the derivation for the on-source region is analogous.

Let
\begin{equation}
	Q_\tot \equiv Q_\on + Q_\off = \sum_{i=1}^{N_\tot} E_i^2.
\end{equation}
Since, for the null hypothesis, $\Expval[U_\on] = \Expval[U_\background]$ and $\Expval[U_\off] = \Expval[U_\background]/\alpha$, we can estimate
\begin{equation}
	\left\{\begin{aligned}
		&\hat{\sigma}^2(U_\on) = \frac{\alpha}{1+\alpha} Q_\tot,\\
		&\hat{\sigma}^2(U_\off) = \frac{1}{1+\alpha} Q_\tot,
	\end{aligned}\right.
\label{eq:sigma_energy}
\end{equation}
giving
\begin{equation}
	\boxed{S_U = \frac{U_\on - \alpha U_\off}{\sqrt{\alpha (Q_\on + Q_\off)}}.}
\end{equation}

\subsection[\texorpdfstring{$\mu_E$}{mu_E}-metric]{\texorpdfstring{$\bm{\mu_E}$}{mu_E}-metric}

Since $\hat\mu_E$ is an intensive exposure-independent quantity, we can define our signal as the difference in the average energy in the on- and off-source regions
\begin{equation}
	\expval{E}_S \equiv \expval{E}_\on - \expval{E}_\off
\end{equation}
with a variance
\begin{equation}
	\sigma^2(\expval{E}_S) = \sigma^2(\expval{E}_\on) + \sigma^2(\expval{E}_\off).
\end{equation}
By the Law of Total Variance,
\begin{equation}
	\sigma^2(\expval{E}_\off) = \Expval\left[\sigma^2(\expval{E}_\off \mid N_\off)\right] + \sigma^2(\Expval[\expval{E}_\off \mid N_\off]),
	\label{eq:eve_law}
\end{equation}
where $\Expval[\cdot|\cdot]$ stands for the conditional expected value and $\sigma^2(\cdot|\cdot)$ for conditional variance. Since
\begin{equation}
	\Expval[\expval{E}_\off \mid N_\off] = \Expval\left[ \frac{1}{N_\off} \sum_{i=1}^{N_\off} E_i \mathrel{\Bigg|} N_\off \right] = \Expval[E] = \mu_E 
\end{equation}
is a constant, we find
\begin{equation}
    \sigma^2(\Expval[\expval{E}_\off \mid N_\off])  = 0. 
\end{equation}
Different from the population mean $\Expval[E] \equiv \mu_E$, the sample mean $\expval{E}_\off \equiv \frac{1}{N_\off} \sum_{i=1}^{N_\off}E_i$, given a fixed $N_\off$, has a variance
\begin{equation}
	\sigma^2(\expval{E}_\off \mid N_\off) = \sigma^2\left( \frac{1}{N_\off} \sum_{i=1}^{N_\off} E_i \mathrel{\Bigg|} N_\off \right) = \frac{\sigma^2(E)}{N_\off}, 
\end{equation}
giving
\begin{equation}
	\Expval\left[\sigma^2(\expval{E}_\off \mid N_\off)\right] = \sigma^2(E) \, \Expval\left[\frac{1}{N_\off}\right]. 
\end{equation}
By \cref{eq:eve_law},
\begin{equation}
    \left\{\begin{aligned} &\sigma^2(\expval{E}_\off) = \sigma^2(E) \Expval\left[\frac{1}{N_\off}\right], \\
    &\sigma^2(\expval{E}_\on) = \sigma^2(E) \Expval\left[\frac{1}{N_\on}\right],    \end{aligned}\right.
\end{equation}
where the derivation for the on-source region is analogous. The population standard deviation $\sigma^2(E)$ can be estimated via
\begin{equation}
  \hat{\sigma}^2(E) = \expval{E^2} - \expval{E}^2 =  \frac{Q_\off}{N_\off} - \left(\frac{U_\off}{N_\off}\right)^2.
\end{equation}

Let $N$ be $N_\on$ or $N_\off$; consider $N \equiv \mu_N + \var{N}$, where $\mu_N\equiv \Expval[N]$ is the population mean and $\var{N}$ is a small variation. The inverse of $N$ is
\begin{equation}
	\frac{1}{N} = \frac{1}{\mu_N + \var{N}} = \frac{1}{\mu_N} \left( 1 + \frac{\var{N}}{\mu_N} \right)^{-1}.
\end{equation}
Performing a binomial expansion assuming $\var{N}/\mu_N \ll 1$, we find
\begin{equation}
	\frac{1}{N} \approx \frac{1}{\mu_N} \left( 1 - \frac{\var{N}}{\mu_N} + \frac{(\var{N})^2}{\mu_N^2} \right).
\end{equation}
The expected value is
\begin{equation}
	\Expval\left[\frac{1}{N}\right] \approx \frac{1}{\mu_N} \left( 1 + \frac{\sigma^2(N)}{\mu_N^2} \right). 
\end{equation}
Since $N$ follows a Poisson distribution, $\sigma^2(N) = \mu_N$, resulting in
\begin{equation}
	\Expval\left[\frac{1}{N}\right] \approx \frac{1}{\mu_N} \left( 1 + \frac{1}{\mu_N} \right).  
\end{equation}
Using the observed sample count $N$ as the maximum likelihood estimator for the theoretical mean $\mu_N$, our signal-to-noise ratio is
\begin{equation}
 	\boxed{S_E \approx \frac{\expval{E}_\on - \expval{E}_\off}{\sqrt{\frac{\sigma^2(E)}{N_\on}\left(1+\frac{1}{N_\on}\right) + \frac{\sigma^2(E)}{N_\off} \left(1+\frac{1}{N_\off}\right) }}.}
\end{equation}
It is worth noting that in the large-$N$ limit, our significance formula reduces to the form of a standard Two-Sample Z-test.

\section{Validity of Gaussian approximations}
\label{app:validity_gauss}

\begin{figure}
    \centering
    \includegraphics[height=6cm]{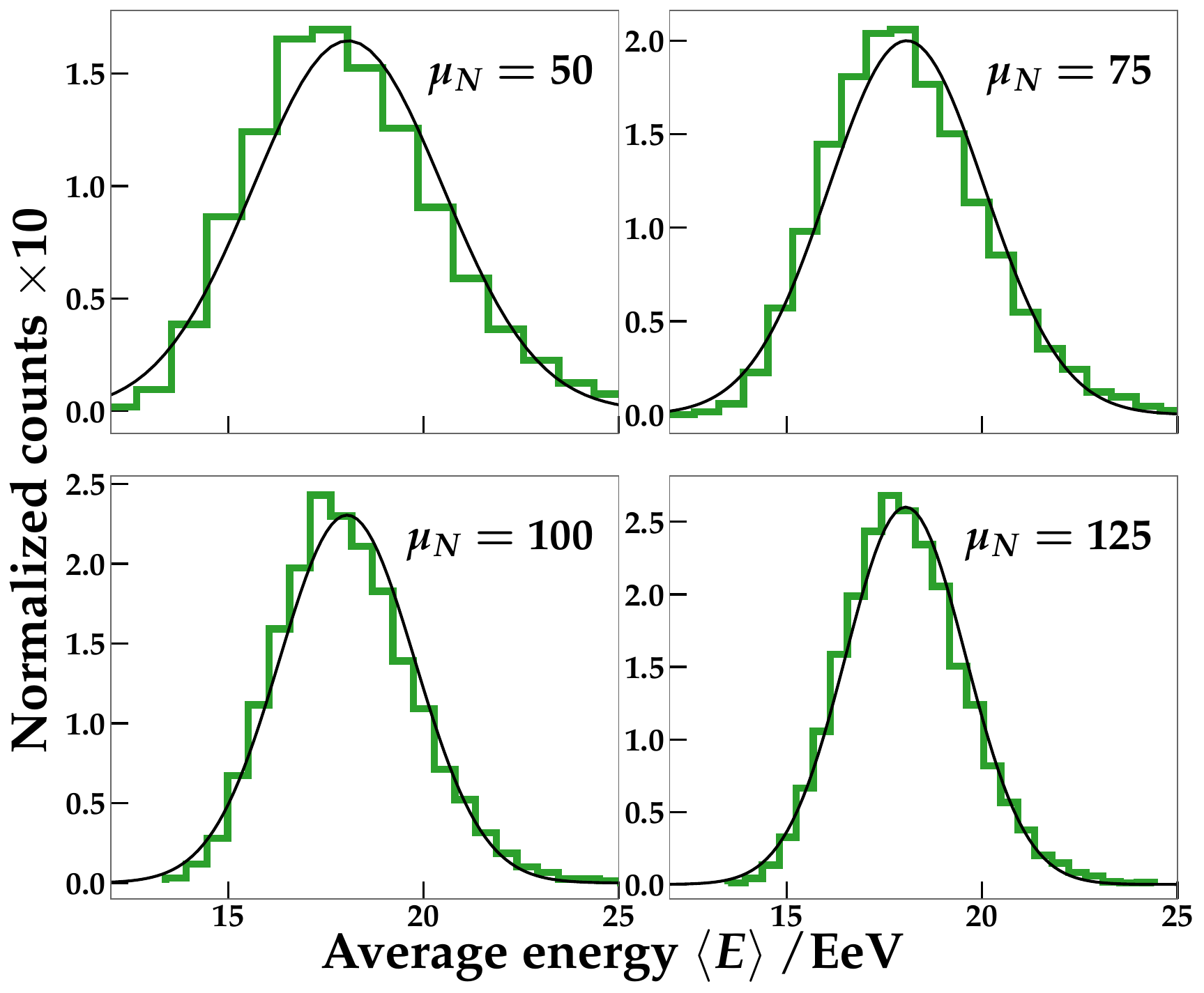}
    \includegraphics[height=6cm]{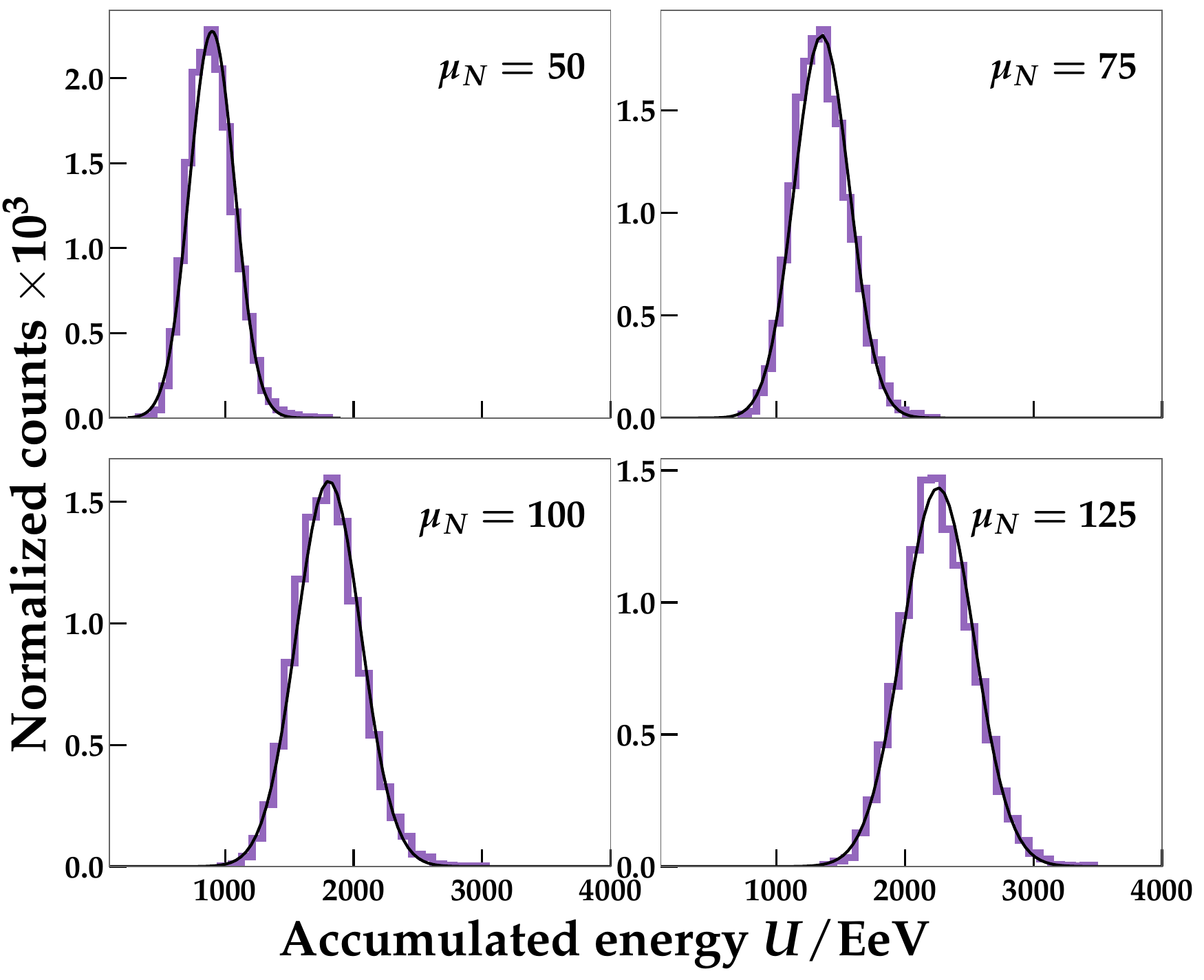}
    \caption{Distribution of $\expval{E}\equiv \frac{1}{N}\sum_{i=1}^{N} E_i$ (left) and $U\equiv \sum_{i=1}^{N} E_i$  (right) for $\num{10000}$ sets of Poisson random numbers $N$ with population mean $\mu_N $. The energy is simulated following a power law with spectral index $\gamma = 2.6$ normalized from 8 to \qty{170}{\EeV}. The black lines show Gaussian distributions fitted to the data. The histograms are normalized such that their integral is equal to 1.}
    \label{fig:gaussian_approximation}
\end{figure}

The validity of the Gaussian approximation done in \cref{app:li-ma} can be verified as follows. We simulate random numbers $N$ following Poisson distributions with population means $\mu_N = 50,75,100,$ and 125. For each random number, we simulate $N$ energies $E$ following a power law with spectral index $\gamma = 2.6$ normalized from 8 to \qty{170}{\EeV}. We show in \cref{fig:gaussian_approximation} the distribution of $\hat\mu_E\equiv \frac{1}{N}\sum_{i=1}^{N} E_i$ (left) and $U\equiv \sum_{i=1}^{N} E_i$ (right) for $\num{10000}$ samples. Even with a small sample size, the distributions can be approximated by Gaussians.

\section{Analytical integration of signal and background}
\label{app:analytical_integration}

In this appendix, we show the procedure used to analytically estimate the number of particles, the accumulated energy, and the average energy within a top-hat region as a function of the threshold energy in \cref{fig:source_and_bg_vs_Eth}.

\subsection[\texorpdfstring{$N$}{N}-metric]{\texorpdfstring{$\bm{N}$}{N}-metric}

Let $f_\source(E)$ be the probability density function associated with the source spectrum, normalized such that $\int_{\Emin}^{\Emax} f_\source(E)\dd{E} = 1$. Analogously, we define $f_\background(E)$ with $\int_{\Emin}^{\Emax} f_\background(E)\dd{E} = 1$ for the background.
The number of particles with energies above $\Eth$ coming from the source is given by
\begin{equation}
    N_\source(\geq\!\Eth) \equiv \sum_{i=1}^{N} I_i
\end{equation}
where $N\equiv N_\source(\geq\!\Emin)$ is the number of injected particles from the source, and $I_i$ is the Bernoulli indicator, defined as
\begin{equation}
    I_i = \begin{cases}
        1, & \text{if $E_i\geq \Eth$},\\
        0, & \text{if $E_i < \Eth$}.
    \end{cases}
\end{equation}
By Wald's Lemma~\cite{wald1944cumulative}, the expected number of particles from the source with energies above $\Eth$ can be calculated using
\begin{equation}
    N_\source(\geq\!\Eth) =  N_\source(\geq\!\Emin) \int_{\Eth}^{\Emax} f_\source(E) \dd{E}.
\end{equation}
The number of background particles with energies above $\Eth$ in a certain top-hat region $\Psi$ around $(\alpha_\ts,\delta_\ts)$ is given by
\begin{equation}
    N_\background(\geq\!\Eth;\delta_\ts,\Psi)  = N_\background(\geq\!\Emin;\delta_\ts,\Psi) \int_{\Eth}^{\Emax} f_\background(E) \dd{E},
\end{equation}
where $N_\background(\geq\!\Emin;\delta,\Psi)$ is the expected number of background particles injected within a top-hat window of radius $\Psi$ centered on the source direction $(\alpha_s, \delta_s)$ under an isotropic sky hypothesis and can be estimated as~\cite{sommers2001cosmic}
\begin{equation}
   \begin{gathered}
       N_\background(\delta_\ts,\Psi;\geq\!E_\tmin) =  N_\background^\text{total}  \frac{\EE_\Psi}{\EE_\text{total}} = N_\background^\text{total}  \frac{ \int_{\delta_\ts-\Psi}^{\delta_\ts+\Psi} \int_{\alpha_\ts-\alpha_\text{lim}(\delta)}^{\alpha_\ts+\alpha_\text{lim}(\delta)}\omega(\delta)\cos(\delta)\dd\alpha\dd\delta }{2\pi \int_{-90^\circ}^{45^\circ} \omega(\delta) \cos(\delta) \dd{\delta} },\\
       \text{where }\alpha_\text{lim}(\delta) \equiv \begin{cases}
    0, & \xi > 1,\\
    \pi, & \xi < 1,\\
    \arccos\xi, &\text{otherwise},
\end{cases} \text{ } \quad \xi = \frac{\cos\Psi - \sin\delta_\ts\sin\delta}{\cos\delta_\ts\cos\delta},
   \end{gathered}
   \label{eq:number_bg}
\end{equation}
and $N_\background^\text{total}$ is the total number of background particles in the sky.
The standard deviations can be estimated as $\sqrt{N_\source}$ and $\sqrt{N_\background}$ for the source and background, respectively.

\subsection[\texorpdfstring{$U$}{U}-metric]{\texorpdfstring{$\bm{U}$}{U}-metric}

The accumulated energy for $E\geq\Eth$ is given by
\begin{equation}
    U(\geq\!\Eth) = \sum_{i=1}^{N} E_i I_i.
\end{equation}
Analogously to the derivation for the $N$-metric, the expected accumulated energy above $\Eth$ is
\begin{equation}
    \left\{ \begin{aligned}
        & U_\source(\geq\!\Eth) =  N_\source(\geq\!\Emin) \int_{\Eth}^{\Emax} E f_\source(E) \dd{E},\\
        & U_\background(\geq\!\Eth;\delta_\ts,\Psi)  = N_\background(\geq\!\Emin;\delta_\ts,\Psi) \int_{\Eth}^{\Emax} E f_\background(E) \dd{E}.
    \end{aligned} \right.
\end{equation}
Due to the Compound-Poisson nature of the distribution, the variances are given by
\begin{equation}
    \left\{ \begin{aligned}
        &\sigma^2_\source(\geq\!\Eth) =  N_\source(\geq\!\Emin) \int_{\Eth}^{\Emax} E^2 f_\source(E) \dd{E},\\
        &\sigma^2_\background(\geq\!\Eth;\delta_\ts,\Psi)  = N_\background(\geq\!\Emin;\delta_\ts,\Psi) \int_{\Eth}^{\Emax} E^2 f_\background(E) \dd{E}.
    \end{aligned}\right.
\end{equation}

\subsection[\texorpdfstring{$\mu_E$}{mu_E}-metric]{\texorpdfstring{$\bm{\mu_E}$}{mu_E}-metric}

The average energy due to the source can be calculated using
\begin{equation}
    \mu_\text{$E$, source}(\geq\!\Eth) =  \frac{\int_{\Eth}^{\Emax} E f_\source(E) \dd{E}}{\int_{\Eth}^{\Emax} f_\source(E) \dd{E}}.
\end{equation}
For the background, we have
\begin{equation}
    \mu_\text{$E$, bg}(\geq\!\Eth)  =  \frac{\int_{\Eth}^{\Emax} E f_\background(E) \dd{E}}{\int_{\Eth}^{\Emax}f_\background(E) \dd{E}}.
\end{equation}
The variances are given by
\begin{equation}
    \left\{ \begin{aligned}
        &\sigma^2_\source(\geq\!\Eth) =  \frac{\int_{\Eth}^{\Emax} E^2 f_\source(E) \dd{E}}{\int_{\Eth}^{\Emax} f_\source(E) \dd{E}} - \mu_\text{$E$, source}^2(\geq\!\Eth),\\
        &\sigma^2_\background(\geq\!\Eth)  = \frac{\int_{\Eth}^{\Emax} E^2 f_\background(E) \dd{E}}{\int_{\Eth}^{\Emax} f_\background(E) \dd{E}} - \mu_\text{$E$, bg}^2(\geq\!\Eth).
    \end{aligned}\right.
\end{equation}

\section{Asymptotic limit of the significance formulae}
\label{app:significance_approximation}

In this appendix, we derive some approximations for the significance formulae $S_N$, $S_U$, and $S_E$ for a weak source, $N_\source\ll N_\background$, and for a small top-hat angle, $\omega_\on\ll \omega_\off$. Although the approximated formulae are not recommended to be used with data, they are useful to understand the qualitative behavior of the significances.

\subsection[\texorpdfstring{$N$}{N}-metric]{\texorpdfstring{$\bm{N}$}{N}-metric}

Apart from the sign, eq.\ (17) from Li--Ma~\cite{LiMa1983} gives the significance formula for the particle excess as
\begin{equation}
    S_N = \sqrt{2} \left[ N_{\on} \ln\left( \underbrace{ \frac{1+\alpha}{\alpha} \frac{N_{\on}}{N_{\on}+N_{\off}}}_{\equiv A} \right) + N_{\off} \ln\left( \underbrace{ (1+\alpha) \frac{N_{\off}}{N_{\on}+N_{\off}}}_{\equiv B} \right) \right]^{1/2}.
\end{equation}
We can rewrite $A$ and $B$ in terms of the number of particles in the background $N_\background = \alpha N_\off$ and in the source $N_\source = N_\on - N_\background$, giving
\begin{equation}
    \left\{\begin{aligned}
        &A  =  \frac{1 + N_\source/N_\background}{1 + \frac{N_\source}{N_\background} \left(\frac{\alpha}{1+\alpha}\right)},\\
        & B = \frac{1}{1 + \frac{N_\source}{N_\background} \left(\frac{\alpha}{1+\alpha}\right)}.
    \end{aligned}\right.
\end{equation}
Under the weak-source assumptions, $N_\source \ll N_\background$, we find 
\begin{equation}
    \frac{1}{1 + \frac{N_\source}{N_\background} \left(\frac{\alpha}{1+\alpha}\right)} \approx 1 - \frac{N_\source}{N_\background} \left(\frac{\alpha}{1+\alpha}\right),
\end{equation}
giving
\begin{equation}
    A \approx 1 + \frac{N_\source/N_\background}{1+\alpha} \quad \text{and} \quad B \approx 1 - \frac{N_\source}{N_\background} \frac{\alpha}{1+\alpha}.
\end{equation}
Since $\ln(1\pm\epsilon) \approx \pm\epsilon - \epsilon^2/2$ for $\epsilon\ll 1$, we find
\begin{equation}
    \ln A \approx \frac{N_\source/N_\background}{1+\alpha} - \frac{(N_\source/N_\background)^2}{2(1+\alpha)^2}.
\end{equation}
Given $N_\on = N_\background(1+N_\source/N_\background)$,
\begin{equation}
    N_\on \ln A \approx N_\background \left[ \frac{N_\source/N_\background}{1+\alpha} + \frac{(N_\source/N_\background)^2}{1+\alpha} - \frac{(N_\source/N_\background)^2}{2(1+\alpha)^2}  \right].
\end{equation}
Analogously,
\begin{equation}
    N_\off \ln B \approx - N_\background \left[ \frac{N_\source/N_\background}{1+\alpha} + \frac{(N_\source/N_\background)^2 \alpha}{2(1+\alpha)^2} \right].
\end{equation}
Summing, we find
\begin{equation}
    S_N \approx \frac{N_\source}{\sqrt{N_\background}} \frac{1}{\sqrt{(1+\alpha)}}.
\end{equation}
Under the assumption $\omega_\off \gg \omega_\on$, we find the signal-to-noise ratio
\begin{equation}
    \boxed{S_N \approx \frac{N_\source}{\sqrt{N_\background}}.}
\end{equation}

\subsection[\texorpdfstring{$U$}{U}-metric]{\texorpdfstring{$\bm{U}$}{U}-metric}
By \cref{eq:LiMa_U},
\begin{equation}
    S_U = U_S/\sqrt{\alpha Q_\tot},
\end{equation}
where $Q_\tot = Q_\on+Q_\off$ and $U_S = U_\on-\alpha U_\off$. The signal is trivially estimated by
\begin{equation}
     U_S = N_\source \hat{\mu}_\text{E, source},
\end{equation}
while the noise is
\begin{equation}
    \sqrt{\alpha Q_\tot} = \sqrt{\alpha N_\tot \expval{E^2}}.
\end{equation}
Under the weak-source assumption, $\expval{E^2} \approx \expval{E^2}_\background$. Hence,
\begin{equation}
    S_U \approx \frac{\hat\mu_\text{$E$, source}}{E_\text{rms, bg}} \left[ \frac{N_\on - \alpha N_\off}{\sqrt{\alpha (N_\on + N_\off)}} \right].
\end{equation}
Identifying the expression under brackets as the Li--Ma significance under the Gaussian limit (eq.\ (9) from Li--Ma~\cite{LiMa1983}),
\begin{equation}
    \boxed{S_U \approx \frac{\hat\mu_\text{$E$, source}}{E_{\text{rms, bg}}} S_N.}
\end{equation}

\subsection[\texorpdfstring{$\mu_E$}{mu_E}-metric]{\texorpdfstring{$\bm{\mu_E}$}{mu_E}-metric}

Under the order $1/N$, \cref{eq:LiMa_E} gives us
\begin{equation}
     S_E = \frac{\hat{\mu}_\text{$E$, on} - \hat{\mu}_\text{$E$, off}}{\sqrt{\frac{\sigma^2(E)}{N_\on} + \frac{\sigma^2(E)}{N_\off}}}.
\end{equation}
The on and off terms can be written as
\begin{equation}
    \left\{\begin{aligned}
        &\hat{\mu}_\text{$E$, on} = \frac{N_\source \hat{\mu}_\text{$E$, source} + N_\background \hat{\mu}_\text{$E$, bg}}{N_\source + N_\background},\\
        &\hat{\mu}_\text{$E$, off} = \hat{\mu}_\text{$E$, bg}.
    \end{aligned}\right.
\end{equation}
Subtracting,
\begin{equation}
    \hat{\mu}_\text{$E$, on} - \hat{\mu}_\text{$E$, off} = \frac{N_\source}{N_\background+N_\source} (\hat{\mu}_\text{$E$, source} - \hat{\mu}_\text{$E$, bg}) = \frac{|N_\on - \alpha N_\off|}{N_\on} (\hat{\mu}_\text{$E$, source} - \hat{\mu}_\text{$E$, bg}).
\end{equation}
Hence,
\begin{equation}
    S_E = \frac{\hat{\mu}_\text{$E$, source} - \hat{\mu}_\text{$E$, bg}}{\sigma(E)} \underbrace{\left| \frac{N_\on - \alpha N_\off}{\sqrt{\alpha (N_\on + N_\off)}} \right|}_{\approx |S_N|} \sqrt{\frac{\alpha N_\off}{N_\on}}.
\end{equation}
Thus,
\begin{equation}
    S_E = \frac{\hat{\mu}_\text{$E$, source} - \hat{\mu}_\text{$E$, bg}}{\sigma(E)} |S_N| \sqrt{\frac{1}{1 + N_\source/N_\background}}.
\end{equation}
Under the weak source assumption, $N_\source \ll N_\background$,
\begin{equation}
    \boxed{S_E \approx \frac{\hat{\mu}_\text{$E$, source} - \hat{\mu}_\text{$E$, bg}}{\sigma(E)} |S_N|.}
\end{equation}

\section{Dispersion across different simulations}
\label{sec:dispersion_simulations}

\begin{figure}
    \centering
    \includegraphics[width=.65\textwidth]{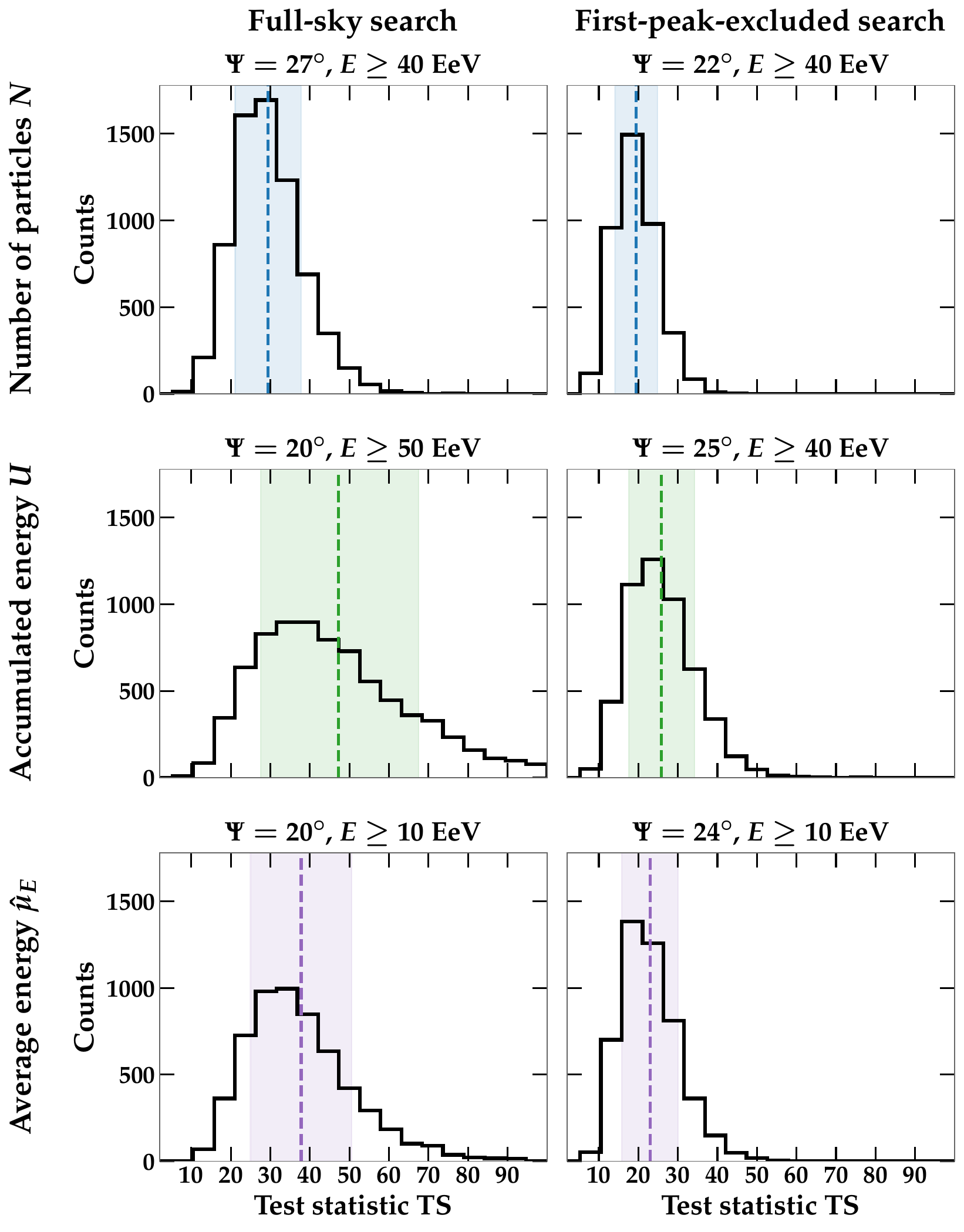}
    \caption{Histogram of the maximum test statistic in the sky map for $\Nsample$ simulations using the $N$ (upper), $U$ (center), and $\hat\mu_E$ (lower) metrics. The left panels correspond to the results obtained when searching the full field of view of the Auger Observatory; the right panels represent those obtained when excluding the sky regions of maximum significance found in the full-sky search. The dashed line and fill correspond to the TS average value and the 68\% confidence level, respectively.}
    \label{fig:hist_all}
\end{figure}

The dispersion of the test statistic TS across $\Nsample$ simulations is presented in \cref{fig:hist_all} for specific threshold energies and top-hat angles. Letting $P_{16}$ and $P_{84}$ represent the 16th and 84th percentiles of the distribution, the lower and upper uncertainties are defined as
\begin{equation}
    \begin{cases}
        \Delta \TS^{-} = \TS - P_{16},\\
        \Delta \TS^{+} = P_{84} - \TS,
    \end{cases}
\end{equation}
where $\TS$ corresponds to the average value of the test statistic evaluated at that energy bin. A mock data set that generates events with extreme energies by chance can inflate or deflate the TS depending on whether such an event happens in the on-source or off-source region (see discussion in \cref{sec:bias_extreme}). The consequence is the production of a few datasets with nonphysical TS, which increases the uncertainties in the $U$ and $\hat\mu_E$-metrics. Overall, the uncertainties in $S$ stay around $1.0\,\sigma$.

\section{Supplementary materials}
\label{app:extra_tables}

In this appendix, we present \cref{tab:global_N,tab:global_U,tab:global_E,tab:local_N,tab:local_U,tab:local_E}.

\begin{table*}
  \centering
  \renewcommand{\arraystretch}{1.1}
  \caption{Results for the $N$-metric across all $E_{\text{th}}$ thresholds for the all-sky search.}
  \label{tab:global_N}
  \resizebox{\textwidth}{!}{%
 \PrintTabGlobalN
  }
\end{table*}

\begin{table*}
  \centering
  \renewcommand{\arraystretch}{1.1}
  \caption{Results for the $N$-metric across all $E_{\text{th}}$ thresholds for the search excluding the first-peak region.}
  \label{tab:local_N}
  \resizebox{\textwidth}{!}{%
 \PrintTabLocalN
  }
\end{table*}

\begin{table*}
  \centering
  \renewcommand{\arraystretch}{1.1}
  \caption{Results for the $U$-metric across all $E_{\text{th}}$ thresholds for the all-sky search.}
  \label{tab:global_U}
  \resizebox{\textwidth}{!}{%
 \PrintTabGlobalU
  }
\end{table*}

\begin{table*}
  \centering
  \renewcommand{\arraystretch}{1.1}
  \caption{Results for the $U$-metric across all $E_{\text{th}}$ thresholds for the search excluding the first-peak region.}
  \label{tab:local_U}
  \resizebox{\textwidth}{!}{%
 \PrintTabLocalU
  }
\end{table*}

\begin{table*}
  \centering
  \renewcommand{\arraystretch}{1.1}
  \caption{Results for the $\hat\mu_E$-metric across all $E_{\text{th}}$ thresholds for the all-sky search.}
  \label{tab:global_E}
  \resizebox{\textwidth}{!}{%
 \PrintTabGlobalE
  }
\end{table*}

\begin{table*}
  \centering
  \renewcommand{\arraystretch}{1.1}
  \caption{Results for the $\hat\mu_E$-metric across all $E_{\text{th}}$ thresholds for the search excluding the first-peak region.}
  \label{tab:local_E}
  \resizebox{\textwidth}{!}{%
 \PrintTabLocalE
  }
\end{table*}

\bibliography{biblio}
\bibliographystyle{JHEP}
\end{document}